\documentclass[floatfix,twocolumn,showpacs,preprintnumbers,amsmath,amssymb,pra,superscriptaddress,longbibliography]{revtex4-1}
\usepackage{color}
\usepackage[usenames,dvipsnames,svgnames,table]{xcolor}
\usepackage[colorlinks=true,linkcolor=blue,urlcolor=blue,citecolor=blue]{hyperref}
\usepackage{mathtools}
\usepackage{graphicx}
\usepackage{dcolumn}
\usepackage{array}
\usepackage{lipsum}
\usepackage{bm}
\usepackage{subfigure}
\usepackage{amssymb}
\usepackage{multirow}
\usepackage{tabularx}
\usepackage{amsmath}
\usepackage{braket}
\usepackage{csquotes}
\graphicspath{{plots/}}
 \usepackage{lipsum}
\usepackage{mathrsfs}

\newcommand{\beq}{\begin{equation}}
\newcommand{\eeq}{\end{equation}}
\newcommand{\bea}{\begin{eqnarray}}
\newcommand{\eea}{\end{eqnarray}}

\begin{document}
\title{
Physical insights from imaginary-time density--density correlation functions
}

\author{Tobias Dornheim}
\email{t.dornheim@hzdr.de}

\affiliation{Center for Advanced Systems Understanding (CASUS), D-02826 G\"orlitz, Germany}
\affiliation{Helmholtz-Zentrum Dresden-Rossendorf (HZDR), D-01328 Dresden, Germany}

\author{Zhandos A.~Moldabekov}

\affiliation{Center for Advanced Systems Understanding (CASUS), D-02826 G\"orlitz, Germany}
\affiliation{Helmholtz-Zentrum Dresden-Rossendorf (HZDR), D-01328 Dresden, Germany}

\author{Panagiotis Tolias}
\affiliation{Space and Plasma Physics, Royal Institute of Technology (KTH), Stockholm, SE-100 44, Sweden}

\author{Maximilian B\"ohme}

\affiliation{Center for Advanced Systems Understanding (CASUS), D-02826 G\"orlitz, Germany}
\affiliation{Helmholtz-Zentrum Dresden-Rossendorf (HZDR), D-01328 Dresden, Germany}
\affiliation{Technische  Universit\"at  Dresden,  D-01062  Dresden,  Germany}

\author{Jan Vorberger}
\affiliation{Helmholtz-Zentrum Dresden-Rossendorf (HZDR), D-01328 Dresden, Germany}

\begin{abstract}
The accurate theoretical description of the dynamic properties of correlated quantum many-body systems such as the dynamic structure factor $S(\mathbf{q},\omega)$ constitutes an important task in many fields. Unfortunately, highly accurate quantum Monte Carlo methods are usually restricted to the imaginary time domain, and the analytic continuation  of the imaginary time density--density correlation function $F(\mathbf{q},\tau)$ to real frequencies is a notoriously hard problem. In this work, we argue that often no such analytic continuation is required as $F(\mathbf{q},\tau)$ contains, by definition, the same physical information as $S(\mathbf{q},\omega)$, only in an unfamiliar representation. Specifically, we show how we can directly extract key information such as the temperature or quasi-particle excitation energies from the $\tau$-domain, which is highly relevant for equation-of-state measurements of matter under extreme conditions [T.~Dornheim \emph{et al.}, \emph{Nature Comm.}~\textbf{13}, 7911 (2022)]. As a practical example, we consider \emph{ab initio} path integral Monte Carlo results for the uniform electron gas (UEG), and demonstrate that even nontrivial processes such as the \emph{roton feature} of the UEG at low density [T.~Dornheim \emph{et al.}, \emph{Comm.~Physics}~\textbf{5}, 304 (2022)] straightforwardly manifest in $F(\mathbf{q},\tau)$. 
In this work, we give a comprehensive overview of various useful properties of $F(\mathbf{q},\tau)$ and how it relates to the usual dynamic structure factor. Moreover, we introduce new concepts such the relation between the frequency moments of $S(\mathbf{q},\omega)$ and the curvature of $F(\mathbf{q},\tau)$ around the origin.
In fact, directly working in the $\tau$-domain is advantageous for many reasons and opens up multiple avenues for future applications.
\end{abstract}
\maketitle

\section{Introduction\label{sec:introduction}}

The accurate theoretical description of nonideal (i.e., interacting) quantum many-body systems is of central importance for a gamut of research fields within physics, quantum chemistry, material science, and related disciplines. While the basic equations governing quantum mechanics have been known for around a century, they are usually too complex to be solved in practice even in the case of a few particles. 
The first attempts to circumvent this obstacle have been based on uncontrolled approximations such as the Hartree-Fock approach, where the difficult treatment of correlations is abandoned in favour of a substantially simplified mean-field picture~\cite{mahan1990many}. Nevertheless, such mean-field methods have given important qualitative insights into a variety of phenomena such as collective plasmon excitations~\cite{pines,bonitz_book,quantum_theory} and Bose-Einstein-condensation~\cite{griffin1996bose,Yukalov2011}.

Over the last decades, the exponential increase in the availability of compute time has sparked a remarkable surge of activity in fields such as computational physics and computational chemistry. In particular, state-of-the-art numerical methods often allow to drastically reduce or even completely avoid approximations and simplifications. In this regard, a case in point is given by the density functional theory (DFT) approach~\cite{Jones_RMP_2015}. More specifically, DFT combines an often sufficient level of accuracy with a manageable computation cost, which arguably makes it the most successful electronic structure method available. Indeed, a sizeable fraction of the world's supercomputing time is being spent on DFT calculations, and the number of DFT based scientific publications has been exponentially growing over the past years~\cite{Pribram-Jones-Review_2015}.
In addition, the computationally even more expensive quantum Monte Carlo (QMC) paradigm~\cite{cep,Foulkes_RMP_2001,anderson2007quantum} is even capable to give exact results in many situations~\cite{Booth2013}, both at finite temperature~\cite{Berne_JCP_1982} and in the ground state.

From a practical perspective, a particularly important property of quantum many-body systems is their response to an external perturbation~\cite{Dornheim_review}. Such \emph{linear-response} properties can be readily measured in scattering experiments~\cite{nolting} and, in principle, give access to the full thermodynamic information of a system. In this context, the key quantity is given by the dynamic structure factor~\cite{quantum_theory}
\begin{eqnarray}\label{eq:DSF}
S(\mathbf{q},\omega) = \int_{-\infty}^\infty \textnormal{d}t\ e^{i\omega t} F(\mathbf{q},t)\ ,
\end{eqnarray}
where the intermediate scattering function at the wave vector $\mathbf{q}$ is defined as the correlation of two density operators in reciprocal space,
\begin{eqnarray}\label{eq:ISF2}
F(\mathbf{q},t) = \braket{\hat n(\mathbf{q},t)\hat n(-\mathbf{q},0)}\ .
\end{eqnarray}
For example, neutron scattering experiments with ultracold helium have given invaluable insights into the collective excitations of superfluids~\cite{griffin1996bose} and normal quantum liquids~\cite{Trigger,Godfrin2012,Dornheim_SciRep_2022}, including the distinct roton feature~\cite{Dornheim_Nature_2022,Kalman_2010} at intermediate wave numbers.
A second important example is given by the diagnostics of warm dense matter (WDM)---an extreme state~\cite{fortov_review,drake2018high,wdm_book} that naturally occurs in astrophysical objects such as giant planet interiors~\cite{Benuzzi_Mounaix_2014}, and is important for technological applications such as inertial confinement fusion~\cite{hu_ICF}. Here X-ray Thomson scattering (XRTS) experiments~\cite{siegfried_review,kraus_xrts} constitute a highly important method of diagnostics and give insights to important system parameters such as the temperature $T$~\cite{Dornheim_T_2022,Dornheim_T2_2022} or the charge state $Z$.

Unfortunately, the theoretical modelling of \emph{dynamic} properties of correlated quantum many-body systems such as $S(\mathbf{q},\omega)$  is notoriously difficult. In practice, DFT cannot give direct access to two-body correlation functions such as Eq.~(\ref{eq:ISF2}), and time-dependent formulations~\cite{marques2012fundamentals} require additional approximations, such as the dynamic exchange--correlation (XC) kernel or a time-dependent XC-potential. 
Furthermore, exact QMC methods are usually restricted to the \emph{imaginary-time} domain. For example, the \emph{ab initio} path integral Monte Carlo (PIMC) approach~\cite{cep,Berne_JCP_1982,Dornheim_permutation_cycles} gives straightforward access to $F(\mathbf{q},\tau)$, corresponding to the intermediate scattering function evaluated at the imaginary time $t=-i\tau$ with $\tau\in[0,\beta]$ and $\beta=1/T$ (we assume Hartree atomic units throughout this work). The connection to the sought-after dynamic structure factor is then given by a two-sided Laplace transform,
\begin{eqnarray}\label{eq:Laplace}
F(\mathbf{q},\tau) &=& \mathcal{L}\left[S(\mathbf{q},\omega)\right] \\\nonumber &=& \int_{-\infty}^\infty \textnormal{d}\omega\ e^{-\omega\tau} S(\mathbf{q},\omega)\ .
\end{eqnarray}
In practice, Eq.~(\ref{eq:Laplace}) constitutes the basis for an \emph{analytic continuation} (AC)~\cite{JARRELL1996133}, i.e., the numerical inversion of $\mathcal{L}\left[S(\mathbf{q},\omega)\right]$ to compute $S(\mathbf{q},\omega)$. This is a well-known and notoriously difficult problem, as any noise in the QMC data for $F(\mathbf{q},\tau)$ leads to instabilities and ambiguity in the DSF. Despite these formidable difficulties, the AC still constitutes one of the most promising routes to rigorously capture the complex interplay of nonideality, quantum degeneracy effects and possibly thermal excitations. Consequently, there exist a multitude of AC methods~\cite{Boninsegni1996,Mishchenko_PRB_2000,Vitali_PRB_2010,Sandvik_PRE_2016,Otsuki_PRE_2017,Goulko_PRB_2017,Boninsegni_maximum_entropy,PhysRevB.98.245101,dornheim_dynamic,dynamic_folgepaper,Fournier_PRL_2020,Nichols_PRE_2022} with different strengths and weaknesses. Unfortunately, one often has to benchmark different methods against each other~\cite{PhysRevB.94.245140}, but the accuracy of the thus reconstructed spectra generally remains unclear.

The present work aims to partially overcome these challenges in the description of the dynamics of correlated quantum many-body systems. More specifically, we argue that, due to the uniqueness of the two-sided Laplace transform, the imaginary-time density--density correlation function (ITCF) $F(\mathbf{q},\tau)$ contains the same information as $S(\mathbf{q},\omega)$ itself, only in a form that might be unfamiliar at the first glance~\cite{Dornheim_PTR_2022}.
While the traditional way of doing physics in the frequency-domain naturally emerges e.g.~from scattering experiments detecting an energy-resolved signal, it is not the only---or necessarily the preferred---option in practice as literally all physical concepts known from the $\omega$-domain
have an analogue in $F(\mathbf{q},\tau)$.
Indeed, many features such as sharp quasi-particle peaks in $S(\mathbf{q},\omega)$ can be identified just as easily in the $\tau$-domain. 
Moreover, we stress that the imaginary-time is directly connected to the physical concept of quantum mechanical delocalization, which means that $F(\mathbf{q},\tau)$ gives straightforward insights that can only be indirectly observed in the DSF. Consequently, it is not our aim to find good approximations to concepts derived in the $\omega$-domain, but instead to highlight how physical effects directly manifest in $F(\mathbf{q},\tau)$.

\begin{figure}\centering
\includegraphics[width=0.5\textwidth]{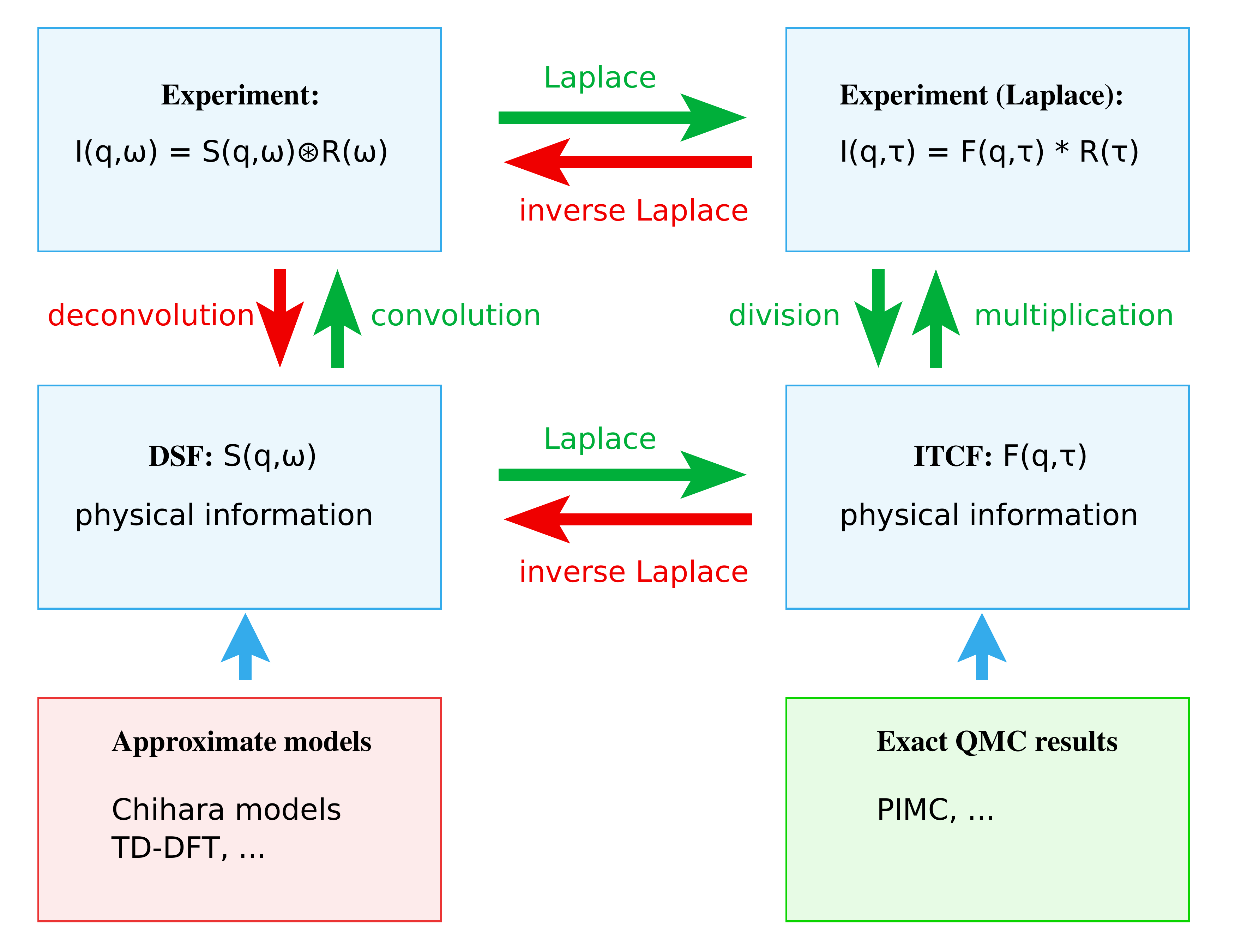}
\caption{\label{fig:sketch}
Advantages of working with $F(\mathbf{q},\tau)$ in the context of an XRTS experiment. Traditionally (left column), an XRTS experiment has been interpreted by constructing an approximate model for the DSF $S(\mathbf{q},\omega)$ and then convolving it with the instrument function $R(\omega)$ to compare with the measured intensity $I(\mathbf{q},\omega)$, cf.~Eq.~(\ref{eq:convolution}); a deconvolution to directly get experimental data for $S(\mathbf{q},\omega)$ is generally not possible. In contrast, working in the $\tau$-domain makes the deconvolution trivial, cf.~Eq.~(\ref{eq:convolution_theorem}), which directly allows to extract important system parameters such as the temperature without any models or simulations~\cite{Dornheim_T_2022}. Furthermore, one can directly compare these experimental results with exact QMC data for $F(\mathbf{q},\tau)$, which opens up the possibility for unprecedented agreement between theory and experiment.
}
\end{figure}

While both domains are formally equivalent, they tend to emphasize different aspects. Therefore, one should analyze both $S(\mathbf{q},\omega)$ and $F(\mathbf{q},\tau)$ when this is possible to get a more complete picture.
In practice, working with $F(\mathbf{q},\tau)$ instead of $S(\mathbf{q},\omega)$ has a number of important benefits that are summarised in Fig.~\ref{fig:sketch} for the case of an XRTS experiment.
In the traditional way, one would construct an approximate model for the DSF, and subsequently convolve it 
with the instrument function $R(\omega)$ that characterises the laser beam and detector to compare to the experimentally measured intensity signal $I(\mathbf{q},\omega)$~\cite{sheffield2010plasma},
\begin{eqnarray}\label{eq:convolution}
I(\mathbf{q},\omega) = S(\mathbf{q},\omega) \circledast R(\omega)\ .
\end{eqnarray}
A straightforward deconvolution of the experimental signal to obtain an experimental $S(\mathbf{q},\omega)$ is generally not possible due to numerical instabilities. 

In stark contrast, these obstacles are completely solved in the $\tau$-domain. Firstly, we note that computing the two-sided Laplace transform of the experimental signal is easy and, in addition, numerically well behaved~\cite{Dornheim_T_2022,Dornheim_T2_2022}. Moreover, the deconvolution is trivial in the $\tau$-domain, as the convolution here works as a simple multiplication; we thus find 
\begin{eqnarray}\label{eq:convolution_theorem}
\mathcal{L}\left[S(\mathbf{q},\omega)\right] = \frac{\mathcal{L}\left[S(\mathbf{q},\omega) \circledast R(\omega)\right]}{\mathcal{L}\left[R(\omega)\right]}\ ,
\end{eqnarray}
where both the enumerator and denominator on the RHS.~of Eq.~(\ref{eq:convolution_theorem}) can be easily evaluated in practice. Therefore, it is possible to directly get experimental results for $F(\mathbf{q},\tau)$, which, in turn give model-free access to physical parameters such as the temperature~\cite{Dornheim_T_2022,Dornheim_T2_2022}. In addition, QMC methods such as PIMC can give exact theoretical results for $F(\mathbf{q},\tau)$, which opens up the enticing possibility for unprecedented agreement between theory and experiment. For completeness, we note that it is also possible to translate well-known approximate models for $S(\mathbf{q},\omega)$ such as the widely used Chihara decomposition~\cite{Chihara_1987} into the $\tau$-domain, where they can be benchmarked against more accurate simulation data.


The paper is organized as follows: in Sec.~\ref{sec:theory}, we introduce the relevant theoretical background starting with the uniform electron gas (UEG) model~\cite{loos,review} that we consider throughout this work; we note, however, that all conclusions that we draw about the physical insights from imaginary-time density--density correlation functions are completely general and in no way particular to the UEG. Sec.~\ref{sec:PIMC} is devoted to the \emph{ab initio} PIMC method~\cite{Berne_JCP_1982,Takahashi_Imada_PIMC_1984,cep} and how it gives straightforward access to $F(\mathbf{q},\tau)$. In addition, we discuss the connection between the DSF and $F(\mathbf{q},\tau)$ to linear-response theory in Sec.~\ref{sec:LRT} followed by a comprehensive overview of some general properties of $F(\mathbf{q},\tau)$ in Sec.~\ref{sec:properties}. In Sec.~\ref{sec:results}, we present our new results, starting with a qualitative investigation of some general trends based on synthetic trial spectra $S(\omega)$ in Sec.~\ref{sec:trends}. This is followed by a detailed investigation of exact PIMC results for $F(\mathbf{q},\tau)$, which give important insights into a number of physical processes including the connection between temperature and imaginary-time diffusion, as well as the nontrivial \emph{roton feature} in the DSF~\cite{Dornheim_Nature_2022}. The paper is concluded by a summary and outlook in Sec.~\ref{sec:summary}.

\section{Theory\label{sec:theory}}

\subsection{The uniform electron gas\label{sec:UEG}}

The UEG~\cite{loos,review} (also known as \emph{jellium} in the literature) is the quantum version of the classical one-component plasma. More specifically, the Hamiltonian of a UEG with $N$ electrons is given by
\begin{eqnarray}\label{eq:Hamiltonian}
\hat{H} = -\frac{1}{2}\sum_{l=1}^N \nabla_l^2 + \sum_{l<k}^N \phi_\textnormal{E}(\hat{\mathbf{r}}_l,\hat{\mathbf{r}}_k) + \frac{N}{2}\xi_\textnormal{M}\ ,
\end{eqnarray}
with $\xi_\textnormal{M}$ being the usual Madelung constant, and the Ewald pair potential $\phi_\textnormal{E}(\hat{\mathbf{r}}_l,\hat{\mathbf{r}}_k)$ taking into account both the interaction between two electrons (and their infinite periodic array of images) and with their respective positive neutralizing homogeneous background. An extensive and accessible discussion of the Ewald potential for the case of the UEG has been given by Fraser \emph{et al.}~\cite{Fraser_PRB_1996}.

From a physical perspective, the UEG can be fully characterised by three reduced parameters~\cite{Ott2018}. 1) the density parameter (also known as Wigner Seitz radius in the literature) $r_s=\overline{r}/a_\textnormal{B}$, with $\overline{r}$ being the average particle separation and $a_\textnormal{B}$ being the first Bohr radius, plays the role of the quantum coupling parameter. Consequently, the UEG attains the limit of a noninteracting Fermi gas for $r_s\to0$ and forms a Wigner crystal for $r_s\gtrsim100$~\cite{Wigner_PhysRev_1934,Jones_Ceperley_PRL_1996,Drummond_PRB_Wigner_2004,Azadi_Wigner_2022}. 2) the degeneracy temperature $\theta=T/T_\textnormal{F}$, with $T_\textnormal{F}$ being the Fermi temperature~\cite{quantum_theory,Ott2018,review}, determines whether the UEG is fully quantum degenerate ($\theta\ll1$) or semi-classical~\cite{Dornheim_HEDP_2022} ($\theta\gg1$). And 3) the spin-polarization parameter $\xi=(N^\uparrow-N^\downarrow)/N$, where $N^\uparrow$ and $N^\downarrow$ are the number of spin-up and spin-down electrons; here we restrict ourselves to the fully unpolarized (i.e., paramagnetic) case of $\xi=0$.
The WDM regime mentioned in the previous section is typically defined by the condition $r_s\sim\theta\sim1$.

We note that the UEG constitutes one of the most fundamental model systems in physics, quantum chemistry, and related fields. Indeed, the accurate parametrization of various UEG properties~\cite{Perdew_Zunger_PRB_1981,Perdew_Wang_PRB_1992,cdop,groth_prl,ksdt,dornheim_ML} based on QMC simulations both in the ground state~\cite{moroni2,Ceperley_Alder_PRL_1980,Spink_PRB_2013} and at finite temperature~\cite{dornheim_prl,Dornheim_PRL_2020,Brown_PRL_2013,Malone_PRL_2016,Schoof_PRL_2015} has been of fundamental importance for a wide spectrum of applications, most notably as input for DFT simulations of real materials.

\subsection{Path integral Monte Carlo and imaginary-time correlation functions\label{sec:PIMC}}

Since its original inception for the description of ultracold $^4$He in the 1960s~\cite{Fosdick_PR_1966}, the \emph{ab initio} PIMC method has become one of the most successful simulation tools in statistical physics and related fields~\cite{Berne_JCP_1982,Takahashi_Imada_PIMC_1984,cep}. The basic idea is to express the canonical partition function (i.e., volume $V$, number density $n=N/V$, and inverse temperature $\beta$ are fixed) in coordinate space, which, for an unpolarized electron gas, gives 
\begin{widetext} 
\begin{eqnarray}\label{eq:Z}
Z_{\beta,N,V} &=& \frac{1}{N^\uparrow! N^\downarrow!} \sum_{\sigma^\uparrow\in S_N^\uparrow} \sum_{\sigma^\downarrow\in S_N^\downarrow} \textnormal{sgn}(\sigma^\uparrow,\sigma^\downarrow) \int_V d\mathbf{R} \bra{\mathbf{R}} e^{-\beta\hat H} \ket{\hat{\pi}_{\sigma^\uparrow}\hat{\pi}_{\sigma^\downarrow}\mathbf{R}}\ .
\end{eqnarray}\end{widetext}
Here the meta variable $\mathbf{R}=(\mathbf{r}_1,\dots,\mathbf{r}_N)^T$ contains the coordinates of all $N=N^\uparrow+N^\downarrow$ electrons in the system.
We note that the anti-symmetry of the fermionic electrons of identical spin-orientation is taken into account by the sums over all possible permutation elements $\sigma^i$ ($i\in\{\uparrow,\downarrow\}$) of the respective permutation group $S_N^i$, and the corresponding permutation operators $\hat{\pi}_{\sigma^i}$. It is easy to see that the density operator $\hat\rho=e^{-\beta\hat{H}}$ in Eq.~(\ref{eq:Z}) can be straightforwardly interpreted as a propagation in imaginary time by an interval of $t=-i\beta$.

Unfortunately, the direct evaluation of Eq.~(\ref{eq:Z}) is precluded by the absent knowledge of the matrix elements $\bra{\mathbf{R}}e^{-\beta\hat{H}}\ket{\mathbf{R}'}$, as the kinetic and potential contributions to the full Hamiltonian $\hat{H}=\hat{K}+\hat{V}$ do not commute. To circumvent this obstacle, we make use of the exact semi-group property 
\begin{eqnarray}\label{eq:group}
e^{-\beta\hat{H}} = \left(e^{-\epsilon\hat{H}} \right)^P \ ,
\end{eqnarray}
with $\epsilon=\beta/P$, leading to
\begin{widetext}
\begin{eqnarray}\label{eq:Z_modified}
Z_{\beta,N,V} &=& \frac{1}{N^\uparrow! N^\downarrow!} \sum_{\sigma^\uparrow\in S_N} \sum_{\sigma^\downarrow\in S_N} \textnormal{sgn}(\sigma^\uparrow,\sigma^\downarrow)\int_V d\mathbf{R}_0\dots d\mathbf{R}_{P-1}
\bra{\mathbf{R}_0}e^{-\epsilon\hat H}\ket{\mathbf{R}_1} \bra{\mathbf{R}_1} \dots 
\bra{\mathbf{R}_{P-1}} e^{-\epsilon\hat H} \ket{\hat{\pi}_{\sigma^\uparrow}\hat{\pi}_{\sigma^\downarrow}\mathbf{R}_0}\ . \quad
\end{eqnarray}
\end{widetext}
Evidently, the single imaginary time propagation has been replaced by $P$ shorter steps of length $\epsilon$. 
From a practical perspective, the main point is that each matrix element has to be evaluated at $P$-times the original temperature. In the limit of large $P$, one can introduce a suitable high-temperature factorization, and the error can be made arbitrarily small.
A detailed discussion of different factorization schemes is beyond the scope of the present work and has been presented in Refs.~\cite{sakkos_JCP_2009,brualla_JCP_2004}.
For the parameters that are of interest for this work, it is sufficient to restrict ourselves to the simple \emph{primitive factorization}~\cite{kleinert2009path}
\begin{eqnarray}\label{eq:primitive}
e^{-\epsilon\hat{H}} = e^{-\epsilon\hat{K}}e^{-\epsilon\hat{V}} + \mathcal{O}\left(\epsilon^{-2}\right)\ .
\end{eqnarray}
We note that the convergence of Eq.~(\ref{eq:primitive}) is ensured by the Trotter formula~\cite{trotter}, and the convergence with $P$ is carefully checked in practice.
Inserting Eq.~(\ref{eq:primitive}) into Eq.~(\ref{eq:Z_modified}) then gives the final expression for the partition function,
\begin{eqnarray}\label{eq:Z_final}
Z &=& \int_V \textnormal{d}\mathbf{X}\ W(\mathbf{X})\\\nonumber
&=& \int_V \textnormal{d}\mathbf{X}\ \prod_{\alpha=0}^{P-1} \left\{
W_V(\mathbf{R}_\alpha) W_K(\mathbf{R}_\alpha,\mathbf{R}_{\alpha+1})
\right\} \ ,
\end{eqnarray}
where the new meta-variable $\mathbf{X}=(\mathbf{R}_0,\dots,\mathbf{R}_{P-1})^T$ contains the coordinates of all particles on all $P$ imaginary-time slices, and it holds $\mathbf{R}_0=\mathbf{R}_P$. For simplicity, we assume that the summation over all permutations, as well as the combinatorial normalisation factors are implicitly contained in $\textnormal{d}\mathbf{X}$.
The weight function $W(\mathbf{X})$ can now be explicitly evaluated, see the second line. More specifically, the potential term is given by
\begin{eqnarray}\label{eq:potential}
W_V(\mathbf{R}_\alpha) = \textnormal{exp}\left(
-\frac{\epsilon}{2}\sum_{k\neq l}^N W(\mathbf{r}_{l,\alpha},\mathbf{r}_{k,\alpha})
\right)\ ,
\end{eqnarray}
with $W(\mathbf{r},\mathbf{s})$ being the pair interaction between two particles [combining both the Ewald pair potential and the Madelung constant from the UEG Hamiltonian Eq.~(\ref{eq:Hamiltonian})], and $\mathbf{r}_{l,\alpha}$ being the coordinate of the l-th particle on imaginary-time slice $\alpha$.
The kinetic term is off-diagonal and corresponds to the density matrix of a noninteracting system,
\begin{eqnarray}\label{eq:kinetic}
W_K(\mathbf{R}_\alpha,\mathbf{R}_{\alpha+1}) =  \frac{\textnormal{exp}\left(-\sum_{l=1}^N\frac{(\mathbf{r}_{l,\alpha}-\mathbf{r}_{l,\alpha+1})^2}{2\sigma_\epsilon^2}\right)}{\left(2\pi\sigma_\epsilon\right)^{3N/2}}\ .\quad
\end{eqnarray}

\begin{figure}\centering
\includegraphics[width=0.475\textwidth]{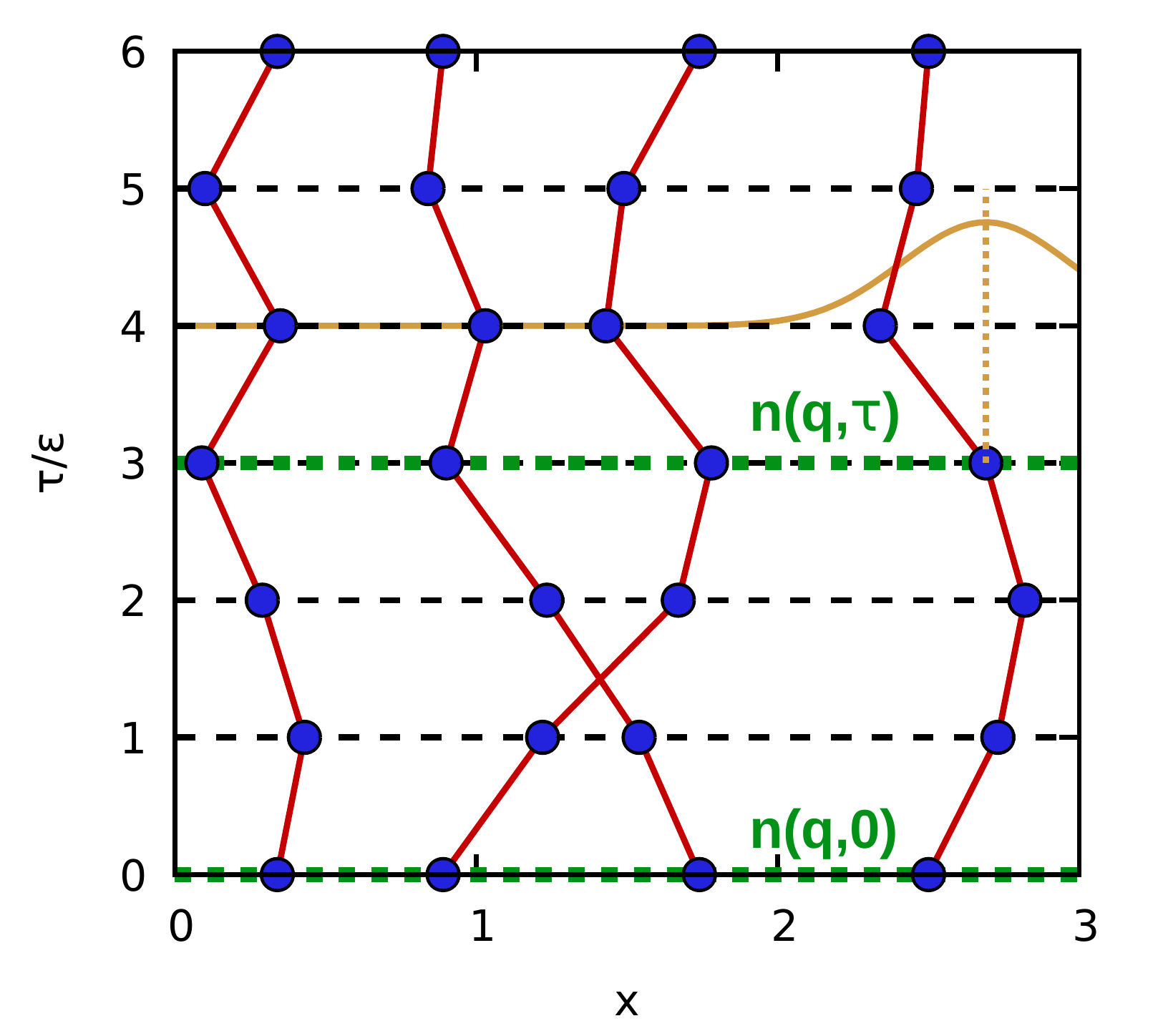}
\caption{\label{fig:Fig}
Illustration of the PIMC method and the corresponding estimation of the imaginary-time density--density correlation function $F(\mathbf{q},\tau)$. Shown is a configuration of $N=4$ particles in the $x$-$\tau$-plane
with $P=6$ imaginary-time propagators. The two horizontal dashed green lines depict the evaluation of a pair of density operators at $\tau=0$ and $\tau=\tau_1$. The yellow Gaussian curve on the right corresponds to the kinetic part of the configuration weight, Eq.~(\ref{eq:kinetic}). Adapted from Ref.~\cite{Dornheim_JCP_ITCF_2021}.
}
\end{figure}

A graphical illustration of Eq.~(\ref{eq:Z_final}) is presented in Fig.~\ref{fig:Fig}, where we show a fictitious configuration of $N=4$ particles in the $\tau$-$x$-plain. First and foremost, we note that each particle is now represented by an entire closed \emph{path} along the imaginary-time direction. This is the well-known classical isomorphism~\cite{Chandler_JCP_1981}, as we have effectively mapped the complicated quantum system of interest onto a system of classical ring polymers. More specifically, the \emph{beads} of each particle interact via the usual pair potential, see Eq.~(\ref{eq:potential}). In addition, beads of the same particle, but on adjacent imaginary-time slices, interact via the harmonic spring potential introduced in Eq.~(\ref{eq:kinetic}). In particular, the characteristic width of the corresponding Gaussian is directly proportional to the thermal wave length $\sigma_\epsilon =\lambda_\epsilon/\sqrt{2\pi}$, with
\begin{eqnarray}\label{eq:lambda}
\lambda_\epsilon = \frac{\lambda_\beta}{\sqrt{P}} \quad \textnormal{and} \quad
\lambda_\beta= \sqrt{2\pi\beta}\ .
\end{eqnarray}
In practice, this spring potential determines the diffusion throughout the imaginary-time, see the yellow Gaussian curve at the RHS.~of Fig.~\ref{fig:Fig}. For high temperature, $\lambda_\beta$ is small and the diffusion is severely restricted. In the classical limit of $\beta\to0$, the Gaussian becomes infinitely narrow, and the paths will collapse to point particles. In contrast, the paths become more extended for low temperatures, which is a direct manifestation of quantum delocalization. The nontrivial interplay of the potential and kinetic terms in Eq.~(\ref{eq:Z_final}) then gives an exact description of quantum diffraction. 
The basic idea of the PIMC method is to randomly generate all possible paths (including all different possible permutations~\cite{Dornheim_permutation_cycles}, see for example the permutation cycle including two identical particles in the center of Fig.~\ref{fig:Fig}) using a particular implementation of the Metropolis algorithm~\cite{metropolis}. In practice, we use an off-diagonal canonical version~\cite{Dornheim_PRB_nk_2021} of the worm algorithm idea by Boninsegni \emph{et al.}~\cite{boninsegni1,boninsegni2} for all calculations in this work.
For completeness, we note that an additional obstacle is given by the antisymmetry of identical fermions under the exchange of particle coordinates [cf.~Eq.~(\ref{eq:Z})], which implies that the configuration weight $W(\mathbf{X})$ can be both positive and negative. This is the origin of the notorious \emph{fermion sign problem}~\cite{troyer,Loh_PRB_1990,dornheim_sign_problem}, which leads to an exponential increase in the compute time with system parameters such as $N$ or $\beta$. While a number of approximate frameworks to deal with the sign problem exist~\cite{Ceperley1991,Brown_PRL_2013,Dornheim_NJP_2015,Yilmaz_JCP_2020,doi:10.1063/5.0030760,Xiong_JCP_2022}, here we carry out exact, unrestricted PIMC simulations. Therefore, our simulations are computationally costly, but exact within the given Monte Carlo error bars. A detailed discussion of the sign problem has been presented in Ref.~\cite{dornheim_sign_problem}.

Let us next consider the estimation of imaginary-time correlation functions within the PIMC paradigm~\cite{Berne_JCP_1983,Boninsegni1996,Dornheim_JCP_ITCF_2021}. In particular, the imaginary-time version of the intermediate scattering function is defined as the imaginary-time density--density correlation function,
\begin{eqnarray}\label{eq:ISF}
F(\mathbf{q},\tau) = \braket{\hat{n}(\mathbf{q},0)\hat{n}(-\mathbf{q},\tau)}\ .
\end{eqnarray}
Naturally, the evaluation of Eq.~(\ref{eq:ISF}) requires the correlated evaluation of the density operator in reciprocal space at two different imaginary-time arguments. Using the factorization introduced in Eq.~(\ref{eq:Z_modified}) it is easy to see that the PIMC method allows for a straightforward evaluation of $F(\mathbf{q},\tau)$ at integer multiples of the factorization step $\epsilon$~\cite{Dornheim_JCP_ITCF_2021},
\begin{widetext}
\begin{eqnarray}\label{eq:F_estimator}
F(\mathbf{q},\tau_j) &=& \frac{1}{Z_{\beta,N,V}} \frac{1}{N^\uparrow! N^\downarrow!} \sum_{\sigma^\uparrow\in S_N^\uparrow} \sum_{\sigma^\downarrow\in S_N^\downarrow} \textnormal{sgn}(\sigma^\uparrow,\sigma^\downarrow) \int d\mathbf{R}_0\dots d\mathbf{R}_{P-1}
\bra{\mathbf{R}_0}{\hat n}(\mathbf{q})e^{-\epsilon\hat H}\ket{\mathbf{R}_1} \bra{\mathbf{R}_1}e^{-\epsilon\hat H}\ket{\mathbf{R}_2}\dots\\\nonumber & & \dots \bra{\mathbf{R}_{j}}{\hat n}(-\mathbf{q})e^{-\epsilon\hat H}\ket{\mathbf{R}_{j+1}} \dots 
\bra{\mathbf{R}_{P-1}} e^{-\epsilon\hat H} \ket{\hat{\pi}_{\sigma^\uparrow}\hat{\pi}_{\sigma^\downarrow}\mathbf{R}_0}\ .
\end{eqnarray}
\end{widetext}
In practice, the $\tau$-grid can be made arbitrarily fine by increasing the number of high-temperature factors $P$, with a linear increase in the required compute time. In contrast, the wave vector-grid is determined by the system size~\cite{Chiesa_PRL_2006,dornheim_prl,dornheim_cpp,Holzmann_PRB_2016}, with $\mathbf{q}=2\pi L^{-1} \mathbf{n}$, and $\mathbf{n}\neq\mathbf{0}$ being an integer vector.

\subsection{Connection to the dynamic structure factor and linear-response theory\label{sec:LRT}}

One of the central relations in the context of the present work is given by Eq.~(\ref{eq:Laplace}), which unambiguously connects the DSF $S(\mathbf{q},\omega)$ to the imaginary-time density--density correlation function $F(\mathbf{q},\tau)$. By definition, both quantities thus contain exactly the same information since the two-sided Laplace transform is a unique transformation.

A second crucial relation is given by the well-known fluctuation--dissipation theorem~\cite{quantum_theory}
\begin{eqnarray}\label{eq:FDT}
S(\mathbf{q},\omega) = - \frac{\textnormal{Im}\chi(\mathbf{q},\omega)}{\pi n (1-e^{-\beta\omega})}\ ,
\end{eqnarray}
which links the DSF to the dynamic density response function $\chi(\mathbf{q},\omega)$ known from linear-response theory. It is convenient to express the latter as~\cite{kugler1},
\begin{eqnarray}\label{eq:LFC}
\chi(\mathbf{q},\omega) = \frac{\chi_0(\mathbf{q},\omega)}{1-\frac{4\pi}{q^2}\left[1-G(\mathbf{q},\omega)\right]\chi_0(\mathbf{q},\omega)}
\end{eqnarray}
with $\chi_0(\mathbf{q},\omega)$ being the analytically known density response function of a noninteracting system, and $G(\mathbf{q},\omega)$ being the dynamic local field correction (LFC) that contains the full information about exchange--correlation effects. Consequently, setting $G(\mathbf{q},\omega)\equiv0$ in Eq.~(\ref{eq:LFC}) leads to a description of the density response on the mean-field level, which is typically known as \emph{random phase approximation} (RPA). 
Evidently, it is straightforward to compute the ITCF from any dielectric theory for $G(\mathbf{q},\omega)$~\cite{stls_original,stls,stls2,tanaka_hnc,castello2021classical,Tolias_JCP_2021,vs_original,schweng,dynamic_ii} by inserting the corresponding (static or dynamic) LFC into Eq.~(\ref{eq:LFC}), evaluating the fluctuation--dissipation theorem Eq.~(\ref{eq:FDT}), and finally computing the two-sided Laplace transform Eq.~(\ref{eq:Laplace}) to obtain $F(\mathbf{q},\tau)$.


Finally, we note that the ITCF is connected to the static limit of Eq.~(\ref{eq:LFC}) via a simple one-dimensional integral,
\begin{eqnarray}\label{eq:chi_static}
\chi(\mathbf{q},0) = - n \int_0^\beta \textnormal{d}\tau\ F(\mathbf{q},\tau)\ ,
\end{eqnarray}
which is sometimes being referred to as imaginary-time version of the fluctuation--dissipation theorem in the literature~\cite{bowen2}, see also the appendix for a short derivation.

\subsection{Properties of the imaginary-time intermediate scattering function\label{sec:properties}}

In this section, we summarise a number of properties of the ITCF. As a starting point, we consider the \emph{detailed balance} relation of the DSF~\cite{siegfried_review,quantum_theory}
\begin{eqnarray}\label{eq:detailed_balance}
S(\mathbf{q},-\omega) = S(\mathbf{q},\omega) e^{-\beta\omega}\ ,
\end{eqnarray}
which holds for uniform systems in thermodynamic equilibrium. Inserting Eq.~(\ref{eq:detailed_balance}) into the two-sided Laplace transform Eq.~(\ref{eq:Laplace}) then directly yields the important symmetry relation~\cite{Dornheim_T_2022}
\begin{eqnarray}\label{eq:symmetry}
F(\mathbf{q},\tau) &=& \int_0^\infty \textnormal{d}\omega\ S(\mathbf{q},\omega)\left\{ e^{-\omega\tau} + e^{-\omega(\beta-\tau)} \right\}\\\nonumber
 &=& F(\mathbf{q},\beta-\tau)\ .
\end{eqnarray}
Clearly, Eq.~(\ref{eq:symmetry}) implies that $F(\mathbf{q},\tau)$ is symmetric around $\tau=\beta/2$. Therefore, any knowledge of the ITCF, say, from an XRTS measurement~\cite{kraus_xrts,siegfried_review,Dornheim_T_2022,Dornheim_T2_2022}, allows one to directly read the temperature from the plot by locating its extremum, which is always a minimum; no theoretical model, simulation, or indirect inference is required.

An additional useful property of a correlated quantum many-body system is given by the frequency moments of the DSF, which we define as
\begin{eqnarray}\label{eq:moments}
M_\alpha = \braket{\omega^\alpha} = \int_{-\infty}^\infty \textnormal{d}\omega\ S(\mathbf{q},\omega)\ \omega^\alpha\ .
\end{eqnarray}
In particular, the odd moments can, in principle, be obtained by evaluating nested commutator terms~\cite{Mihara_Puff_PR_1968}. In the quantum mechanical case~\cite{kugler1}, the cases of $\alpha=-1,1,3$ are known from such sum rules. For example, the well-known f-sum rule gives the important relation~\cite{quantum_theory}
\begin{eqnarray}\label{eq:f_sum_rule}
M_1 = \braket{\omega^1} = \frac{\mathbf{q}^2}{2}\ .
\end{eqnarray}
To connect Eq.~(\ref{eq:moments}) to the ITCF, we may differentiate the latter with respect to $\tau$, which gives
\beq
\frac{d^n}{d\tau^n}F(q,\tau)=(-1)^n\int\limits_{-\infty}^{\infty} d\omega\, \omega^n e^{-\tau\omega}S(q,\omega)\;.
\eeq
Therefore, the derivatives of $F(\mathbf{q},\tau)$ at the origin give us straightforward access to both odd and even frequency moments~\cite{Sandvik_PRB_1998,Dornheim_moments_2023}
\begin{eqnarray}\label{eq:moments_derivative}
M_\alpha = \left( -1 \right)^\alpha \left. \frac{\partial^\alpha}{\partial\tau^\alpha} F(\mathbf{q},\tau) \right|_{\tau=0} \ ,
\end{eqnarray}
for $\alpha\geq 0$. In practice, knowledge of such frequency moments constitutes important input to further constrain a potential AC from the imaginary-time domain to $S(\mathbf{q},\omega)$, see, e.g., Refs.~\cite{Filinov_PRA_2012,Filinov_PRA_2016}. In fact, the DSF is fully determined by the $M_\alpha$, which is known as the Hamburger problem in the respective literature~\cite{tkachenko_book}. Therefore, the frequency moments (and, thus, the derivatives of $F(\mathbf{q},\tau)$ around the origin) contain important physical insights, as we will discuss in Sec.~\ref{sec:results}.

As a final useful relation, we consider the exact spectral representation of the DSF, which is given by~\cite{quantum_theory}
\begin{eqnarray}\label{eq:spectral}
S(\mathbf{q},\omega) = \sum_{m,l} P_m \left\|{n}_{ml}(\mathbf{q}) \right\|^2 \delta(\omega - \omega_{lm})\ .
\end{eqnarray}
Here $l$ and $m$ denote the eigenstates of the Hamiltonian, $\omega_{lm}=(E_l-E_m)/\hbar$ is the corresponding energy difference, and $n_{ml}$ is the transition element from state $m$ to $l$ due to the density operator $\hat{n}(\mathbf{q})$. In addition, $P_m=e^{-E_m\beta}/Z$ is the probability to occupy the initial state $m$.
The two-sided Laplace transform of Eq.~(\ref{eq:spectral}) then gives an analogous spectral representation of the ITCF,
\begin{eqnarray}\label{eq:spectral_F}
F(\mathbf{q},\tau) &=& \sum_{m,l} P_m \left\|{n}_{ml}(\mathbf{q}) \right\|^2 e^{-\tau\omega_{lm}} \\\nonumber
&=& \sum_{E_m<E_l} \left\|{n}_{ml}(\mathbf{q}) \right\|^2 \left\{ 
P_m e^{-\tau\omega_{lm}} + P_l e^{\tau\omega_{lm}}
\right\}\ ,
\end{eqnarray}
where we have ordered the eigenstates according to their energy in the second line. At finite temperature (i.e., $T>0$, leading to a finite value of $\beta$), transitions between eigenstates occur in both directions. Specifically, excitations that increase the energy always lead to an exponential decay with $\tau$ (see also Appendix D of Ref.~\cite{martin2016interacting}), whereas, conversely, excitations that reduce the energy lead to a corresponding exponential increase. We note that this is directly reflected in the symmetry relation Eq.~(\ref{eq:symmetry}).

\section{Results\label{sec:results}}

\subsection{General trends: synthetic spectra\label{sec:trends}}

To get a general feeling for the relation between $S(\mathbf{q},\omega)$ and $F(\mathbf{q},\tau)$, we start by considering synthetic DSFs that are of a simple Gaussian form, but obey the detailed balance relation Eq.~(\ref{eq:detailed_balance}),
\begin{eqnarray}
S(\omega) =  
\begin{cases}
    G(\omega,\omega_0,\sigma),& \text{if } \omega\geq 0\\
    e^{\beta\omega} G(-\omega,\omega_0,\sigma),              & \text{otherwise}\ ,
\end{cases}
\end{eqnarray}
with the usual definition of a normalised Gaussian
\begin{eqnarray}\label{eq:Gauss}
G(\omega,\omega_0,\sigma) = \frac{\textnormal{exp}\left(\frac{(\omega-\omega_0)^2}{2\sigma^2}\right)}{\sqrt{2\pi\sigma^2}}\ .
\end{eqnarray}
The corresponding two-sided Laplace transform Eq.~(\ref{eq:Laplace}) is given by
\begin{eqnarray}\label{eq:F_Gauss}
F(\tau) = A(\tau,\omega_0,\sigma) + A(\beta-\tau,\omega_0,\sigma) \ ,
\end{eqnarray}
with the definition
\begin{eqnarray}
A(\tau,\omega_0,\sigma) = e^{-\tau\omega_0}e^{\tau^2\sigma^2/2}\left(\frac{1+\textnormal{erf}\left(\frac{\omega_0-\sigma^2\tau}{\sqrt{2\sigma^2}}\right)}{2}\right)\ .\quad\quad
\end{eqnarray}

\begin{figure}\centering
\includegraphics[width=0.45\textwidth]{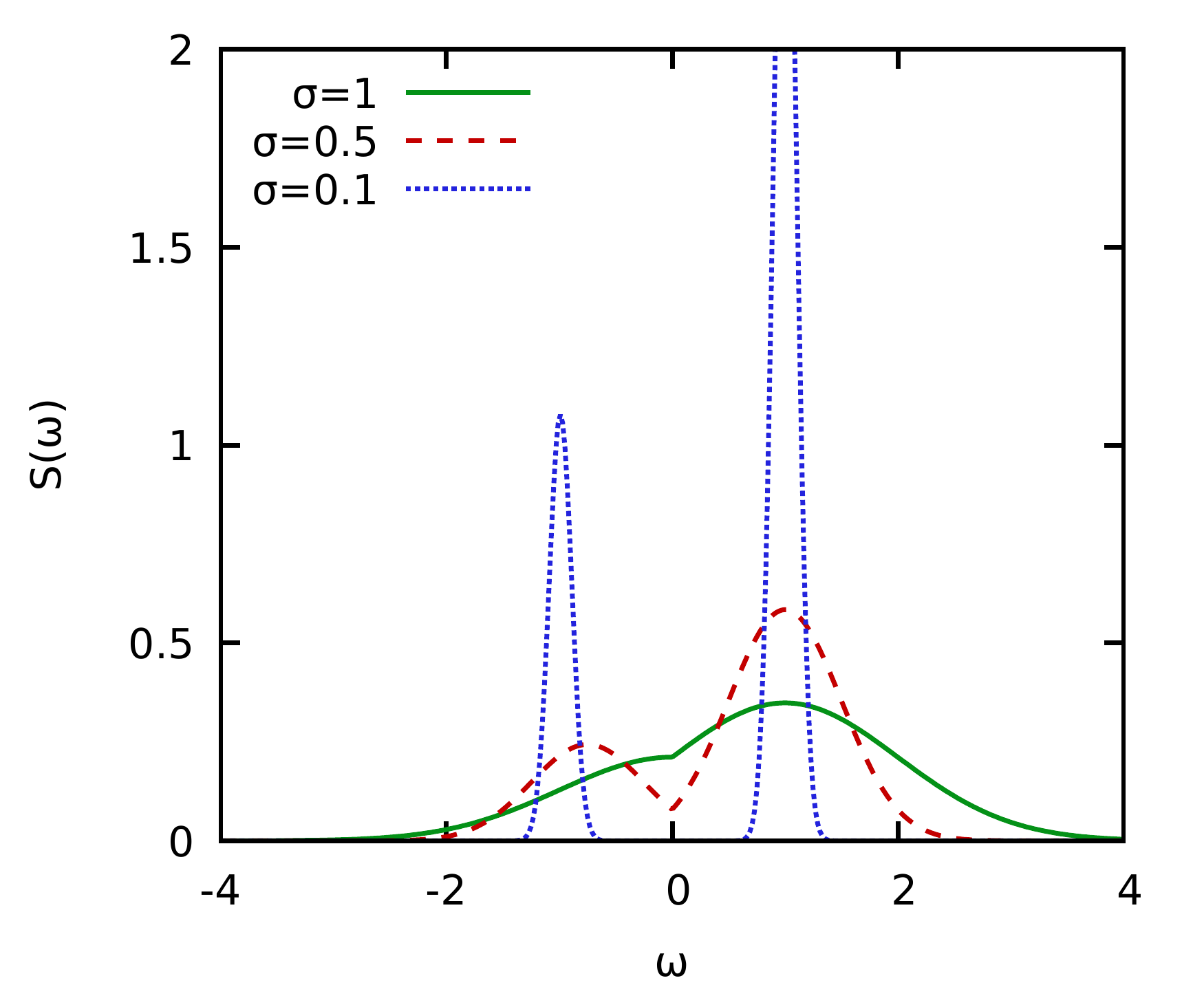}\\
\includegraphics[width=0.45\textwidth]{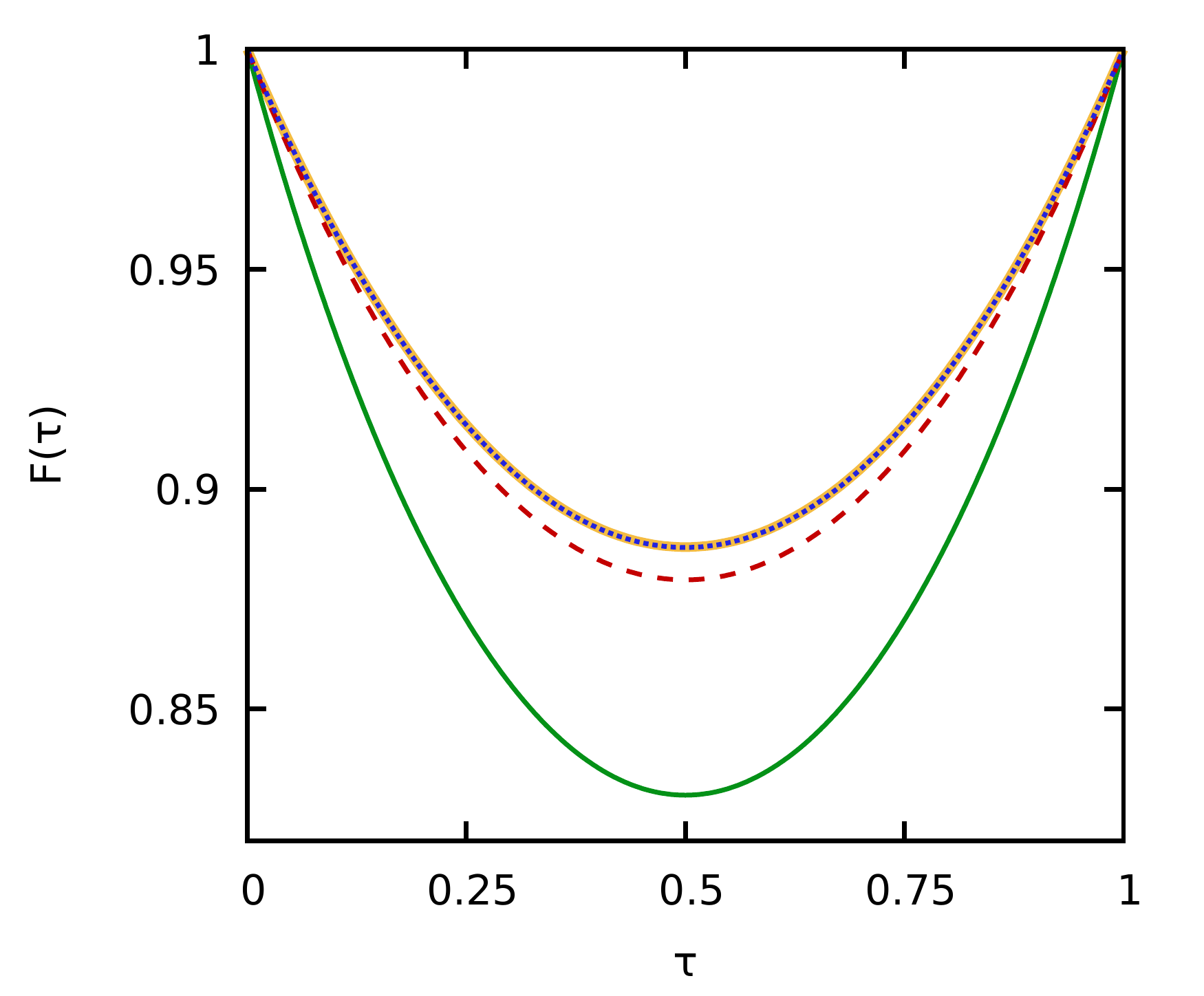}
\caption{\label{fig:Gauss_width}
Top: synthetic Gaussian [Eq.~(\ref{eq:Gauss}] DSFs $S(\omega)$ with $\beta=1$, $\omega_0=1$ and different variance $\sigma$; the units of all properties are arbitrary. Bottom: corresponding two-sided Laplace transform $F(\tau)$, Eq.~(\ref{eq:F_Gauss}). Note that we have normalised both $S(\omega)$ and $F(\tau)$ to $F(0)$.
}
\end{figure}

Let us start the investigation of such synthetic spectra by considering the impact of the peak width $\sigma$. The results are shown in Fig.~\ref{fig:Gauss_width} and the top (bottom) panel shows the results for the DSF with $\beta=1$ and $\omega_0=1$ (the corresponding $F(\tau)$). For $\sigma=1$, the DSF consists of a single broad peak combining the positive and negative frequency range; we note that the damping for $\omega<0$ is a quantum effect described by the detailed balance relation Eq.~(\ref{eq:detailed_balance}). This manifests as a more pronounced decay with $\tau$ in the ITCF compared to the other depicted examples. For $\sigma=0.5$, there remains some significant overlap in $S(\omega)$ around $\omega=0$, but we overall find a double peak structure in the DSF. This pronounced change in the DSF is directly translated to $F(\tau)$ as a less steep decay with $\tau$. Finally, the dotted blue curve shows results for $\sigma=0.1$, which leads to two narrow peaks in $S(\omega)$. In practice, such features are often interpreted as distinct quasi-particle excitations. Indeed, our synthetic model approaches a delta-like quasi-particle (QP) excitation in the limit of $\sigma\to0$
\begin{eqnarray}\label{eq:S_QP}
S_\textnormal{QP}(\omega) = \delta(\omega-\omega_\textnormal{QP}) + e^{-\beta\omega_\textnormal{QP}}\delta(\omega+\omega_\textnormal{QP})\ ,
\end{eqnarray}
with $\omega_\textnormal{QP}$ being the quasi-particle energy, as Eq.~(\ref{eq:Gauss}) is a nascent delta-function. 
It is easy to see that the corresponding ITCF is given by
\begin{eqnarray}\label{eq:F_QP}
F_\textnormal{QP}(\tau) = e^{-\tau\omega_\textnormal{QP}} + e^{-(\beta-\tau)\omega_\textnormal{QP}}\ .
\end{eqnarray}
The results for Eq.~(\ref{eq:F_QP}) are shown as the solid yellow curve in the bottom panel of Fig.~\ref{fig:Gauss_width}, and are in excellent agreement to the ITCF of the Gaussian trial DSF with $\sigma=0.1$. This finding has important consequences for the practical interpretation of scattering experiments. For example, it is well known that XRTS signals of a WDM probe exhibit a sharp plasmon excitation at the plasma frequency $\omega_\textnormal{p}$ for small wave numbers $q=|\mathbf{q}|$.
In this way, the location of the plasmon gives direct insight into the density of the unbound electrons that take part in this collective excitation, and, therefore, constitutes an import tool for diagnostics. In practice, this endeavour is complicated by the convolution with the instrument function, which, in combination with other features in the DSF, might mask the true location of the plasmon.
In contrast, Eq.~(\ref{eq:F_QP}) implies that we can extract the plasma frequency directly from an exponential fit to $F(\mathbf{q},\tau)$. As explained in Fig.~\ref{fig:sketch} above, the latter can be computed without the bias due to the instrument function as the convolution is just a multiplication in the $\tau$-domain. 
We thus conclude that working with $F(\mathbf{q},\tau)$ is not only advantageous to diagnose the temperature of a sample (see the recent work by Dornheim \emph{et al.}~\cite{Dornheim_T_2022}, and the discussion of Fig.~\ref{fig:Gauss_beta} below), but gives also direct access to the free electronic density in a forward-scattering experiment where the scattering angle, and thus the wave number, are small enough to probe the regime of collective plasma excitations.

For completeness, we note that other concepts such as approximate expressions for the plasmon location at finite wave numbers~\cite{Thiele_PRE_2008} that are sometimes used to diagnose plasma parameters~\cite{Glenzer_PRL_2007,Preston_APL_2019,siegfried_review} can also be directly translated into the $\tau$-domain. In general, however, we argue against the blind transformation of $\omega$-native concepts to $F(\mathbf{q},\tau)$, as it makes more sense to directly work with the rich physics that are naturally inherent to $F(\mathbf{q},\tau)$.

\begin{figure}\centering
\includegraphics[width=0.45\textwidth]{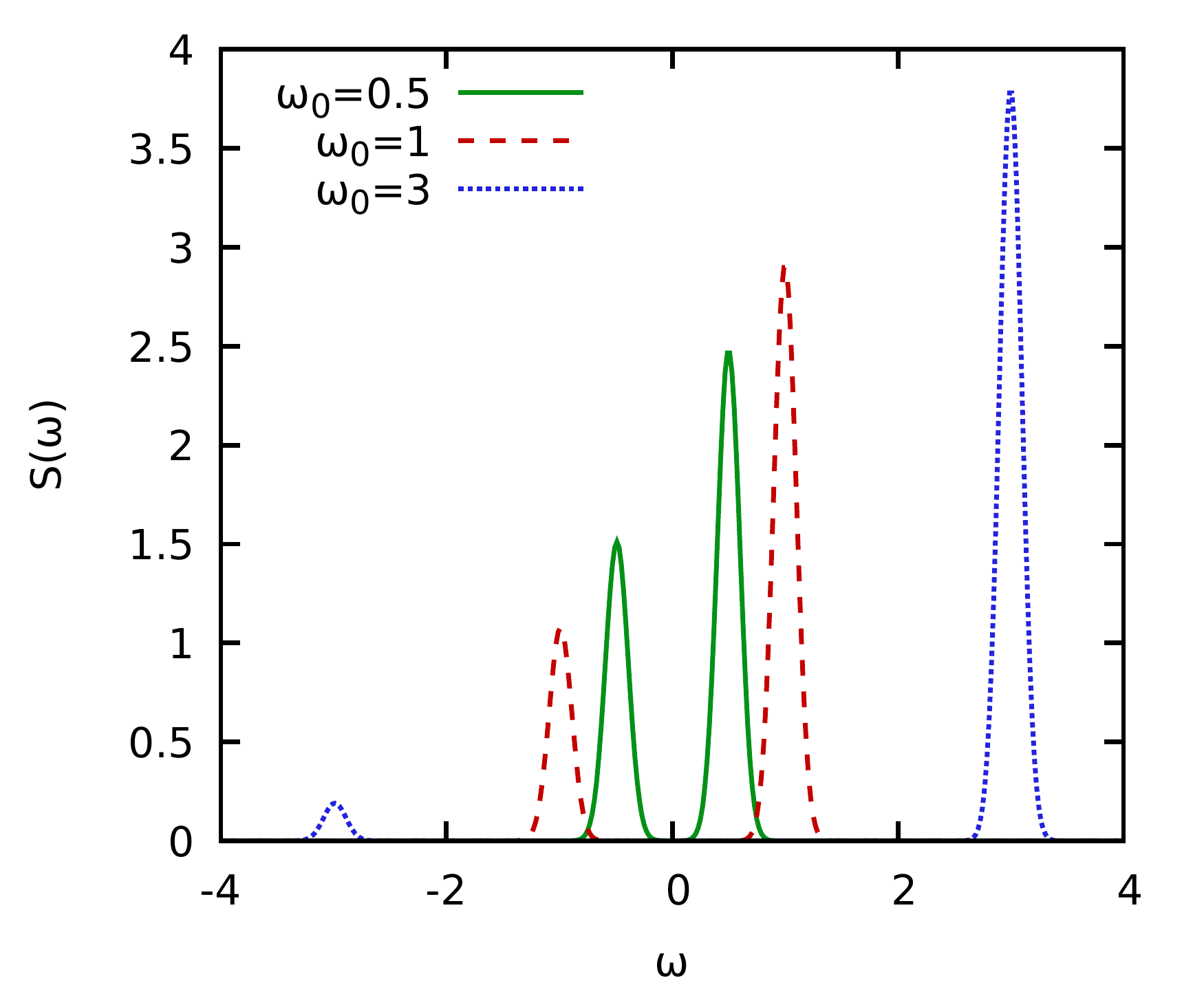}\\
\includegraphics[width=0.45\textwidth]{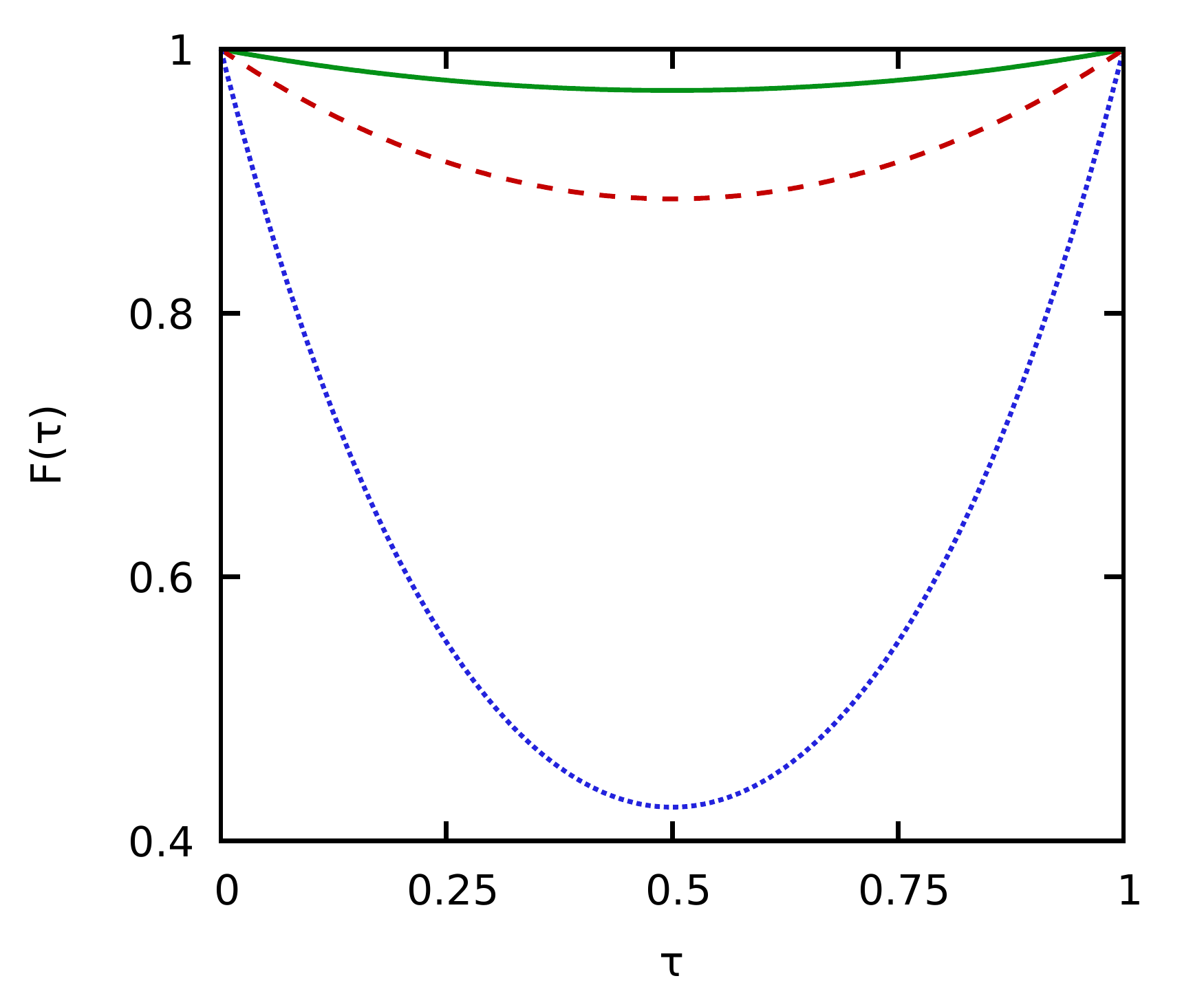}
\caption{\label{fig:Gauss_w0}
Top: synthetic Gaussian [Eq.~(\ref{eq:Gauss}] DSFs $S(\omega)$ with $\beta=1$, $\sigma=0.1$, and different values of $\omega_0$; the units of all properties are arbitrary. Bottom: corresponding two-sided Laplace transform $F(\tau)$, Eq.~(\ref{eq:F_Gauss}). Note that we have normalised both $S(\omega)$ and $F(\tau)$ to $F(0)$.
}
\end{figure}

\begin{figure}\centering
\includegraphics[width=0.45\textwidth]{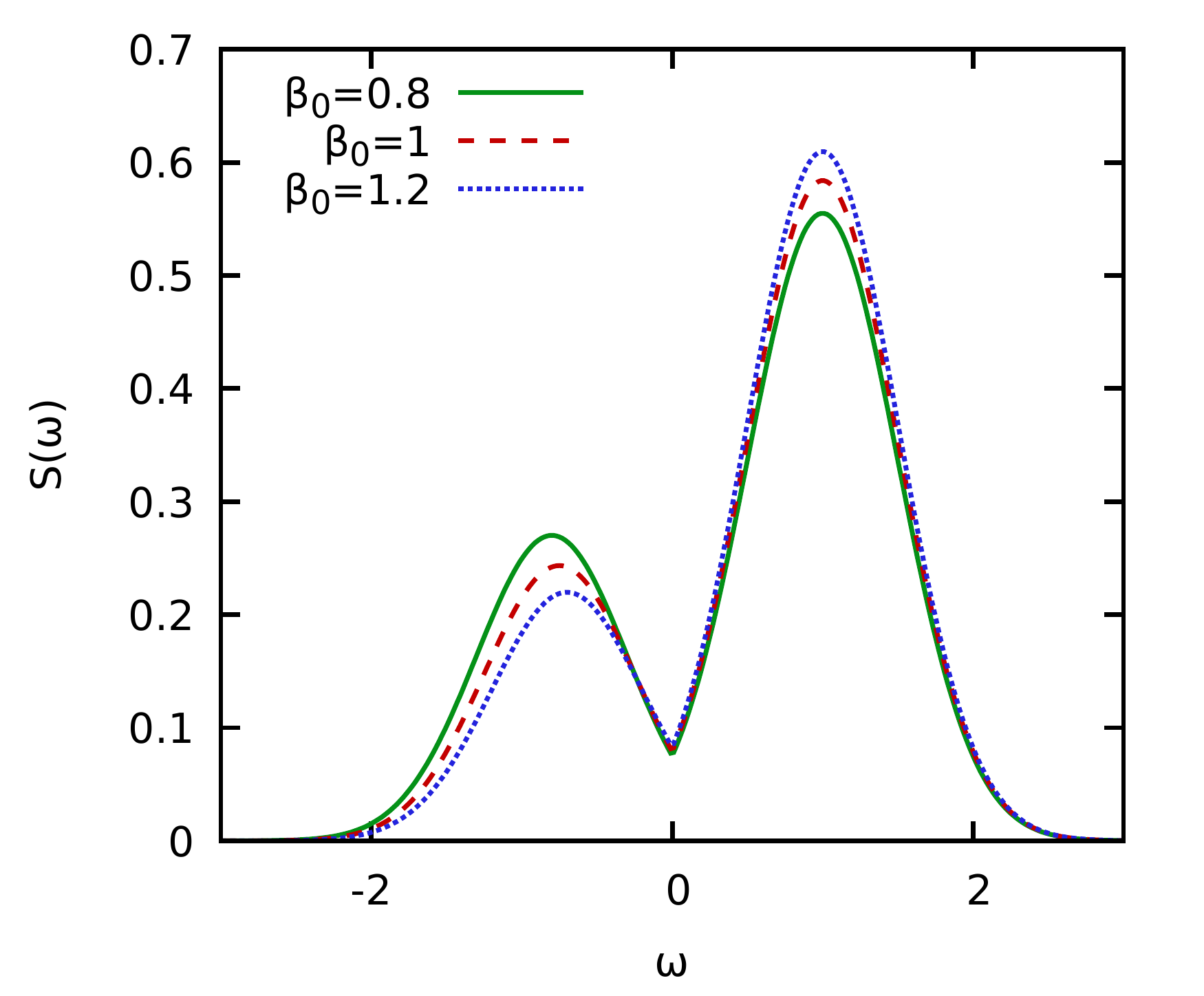}\\
\includegraphics[width=0.45\textwidth]{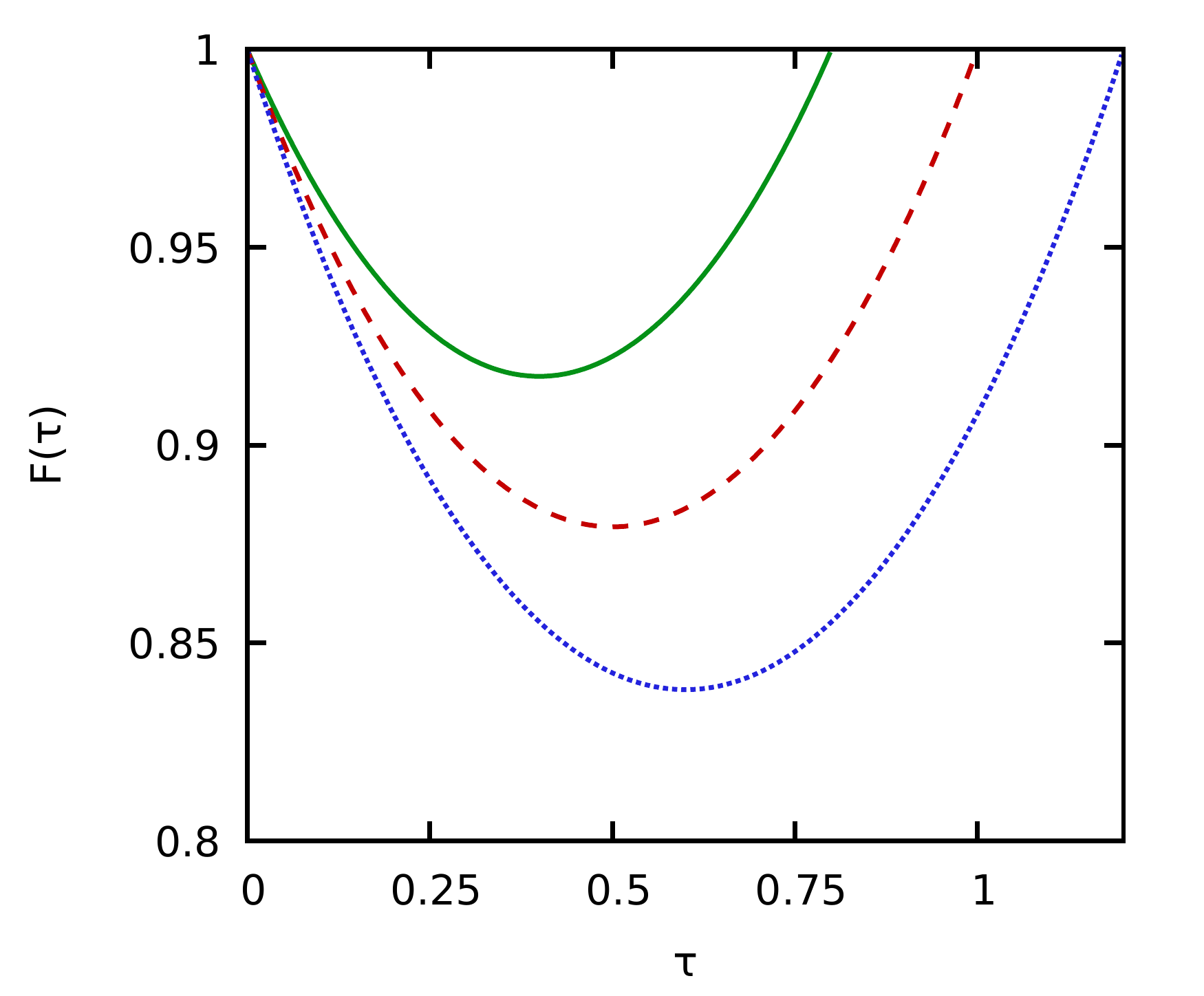}
\caption{\label{fig:Gauss_beta}
Top: synthetic Gaussian [Eq.~(\ref{eq:Gauss}] DSFs $S(\omega)$ with $\omega_0=1$, $\sigma=0.5$, and slightly different values of the inverse temperature $\beta$; the units of all properties are arbitrary. Bottom: corresponding two-sided Laplace transform $F(\tau)$, Eq.~(\ref{eq:F_Gauss}). Note that we have normalised both $S(\omega)$ and $F(\tau)$ to $F(0)$.
}
\end{figure} 

Let us continue our investigation of synthetic spectra by investigating the manifestation of the peak position $\omega_0$ shown in Fig.~\ref{fig:Gauss_w0}. In fact, the impact of the peak position can qualitatively be immediately seen from Eqs.~(\ref{eq:spectral_F}) and (\ref{eq:F_QP}). Specifically, transferring spectral weight from larger towards smaller excitation energies $\omega$ leads to a less steep decay along $\tau$ (for $\tau<\beta/2$). This is indeed what we find in the bottom panel of Fig.~\ref{fig:Gauss_w0}.

From a physical perspective, this finding has important implications, as we shall see in the discussion of real physical results for the UEG in Sec.~\ref{sec:UEG} below. Firstly, it means that the persistence of two-body correlations throughout the imaginary time $0\leq\tau\leq\beta/2$, which we can directly observe in our PIMC simulations, implies a down-shift in the dominating excitation energies in $S(\mathbf{q},\omega)$, see the spectral representation Eq.~(\ref{eq:spectral}) above. This is strongly related to the \emph{roton feature} in the spectrum of density fluctuations both in the UEG~\cite{Dornheim_Nature_2022}, but also quantum liquids such as ultracold helium~\cite{griffin1996bose,Godfrin2012,Dornheim_SciRep_2022,Trigger}. 
In other words, quantifying the decay of correlations in $F(\mathbf{q},\tau)$ constitutes a straightforward alternative to the usual pseudo dispersion relation $\omega(q)$ constructed from the position of the maximum in the DSF~\cite{Hamann_CPP_2020}.
In addition, the aforementioned effects lead to a maximum in the static linear density response function $\chi(\mathbf{q},0)$. This can be seen either from the imaginary-time version of the fluctuation--dissipation theorem Eq.~(\ref{eq:chi_static}), or by re-calling the relation between $\chi(\mathbf{q},0)$ and the inverse frequency-moment of the DSF~\cite{dornheim_dynamic,dynamic_folgepaper},
\begin{eqnarray}
M_{-1} = - \frac{\chi(\mathbf{q},0)}{2n}\ .
\end{eqnarray}

\begin{figure*}\centering
\includegraphics[width=0.475\textwidth]{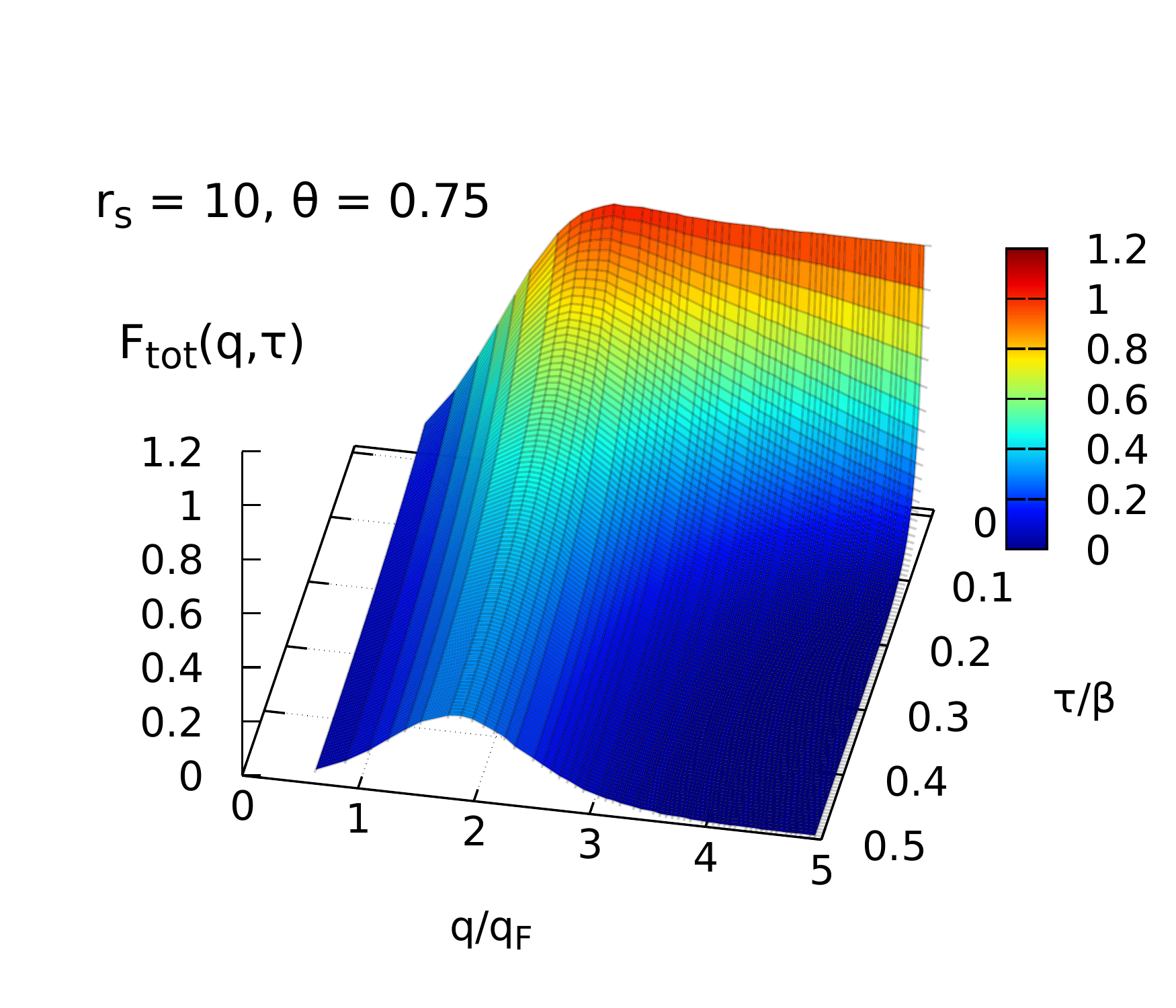}\includegraphics[width=0.475\textwidth]{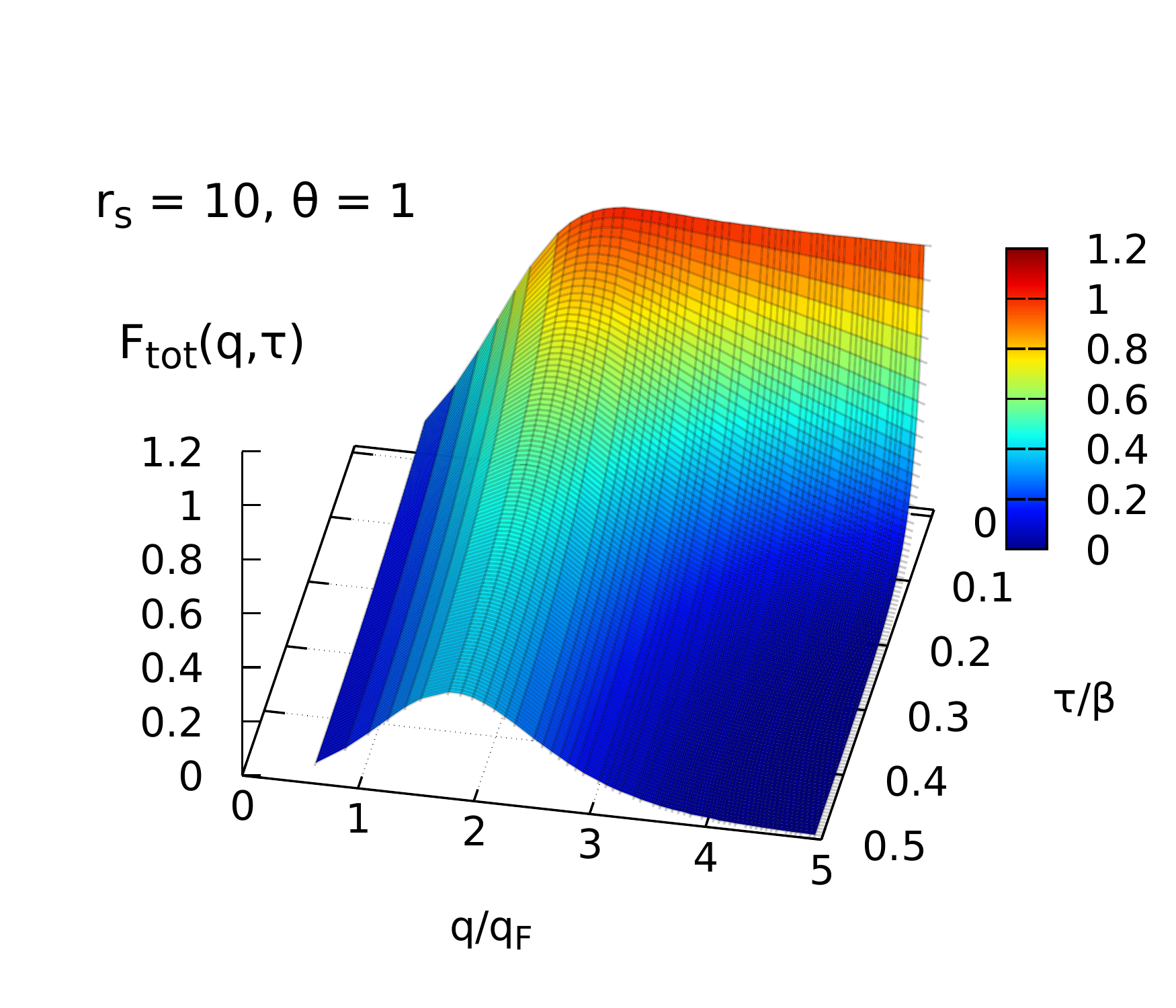}\\\vspace*{-1.4cm}
\includegraphics[width=0.475\textwidth]{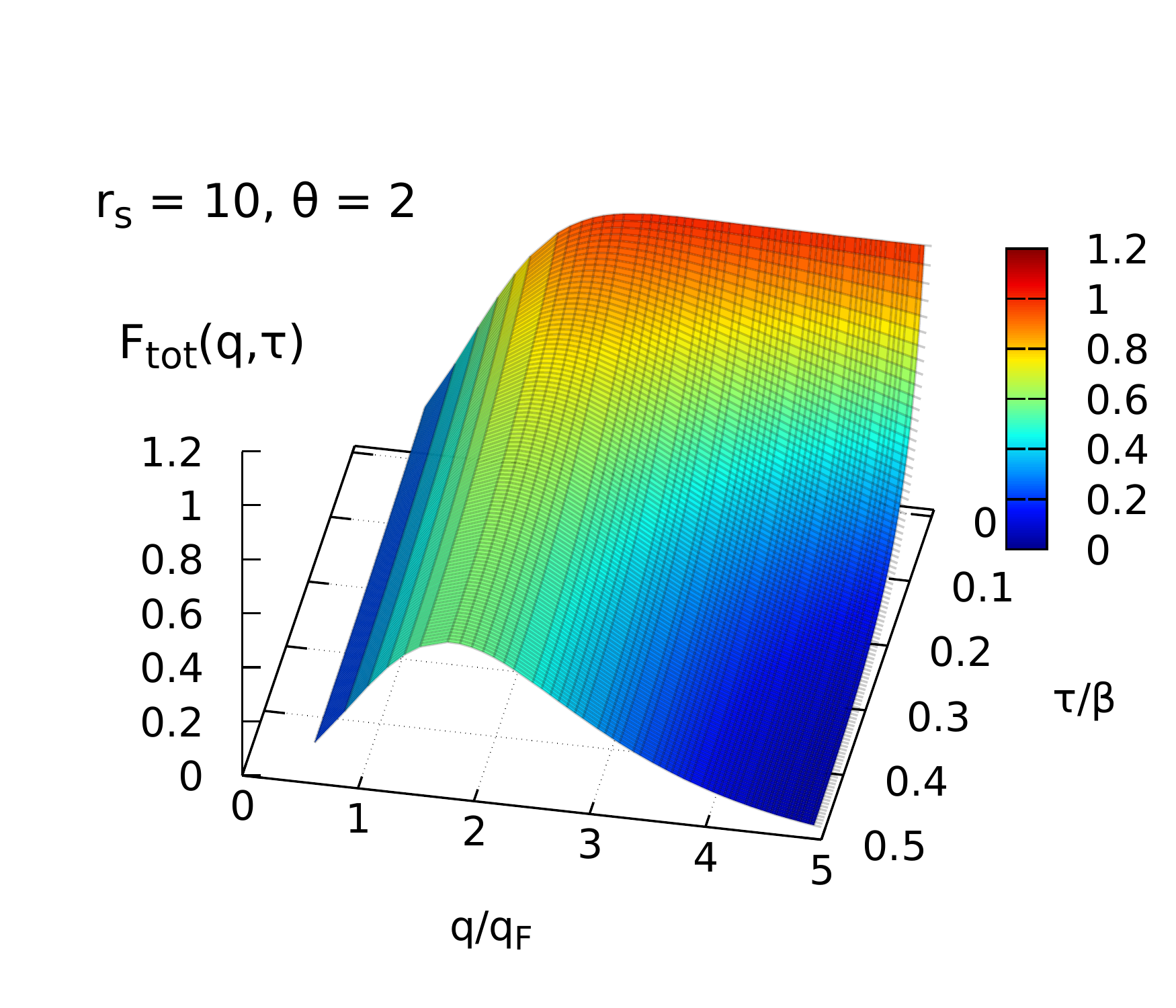}\includegraphics[width=0.475\textwidth]{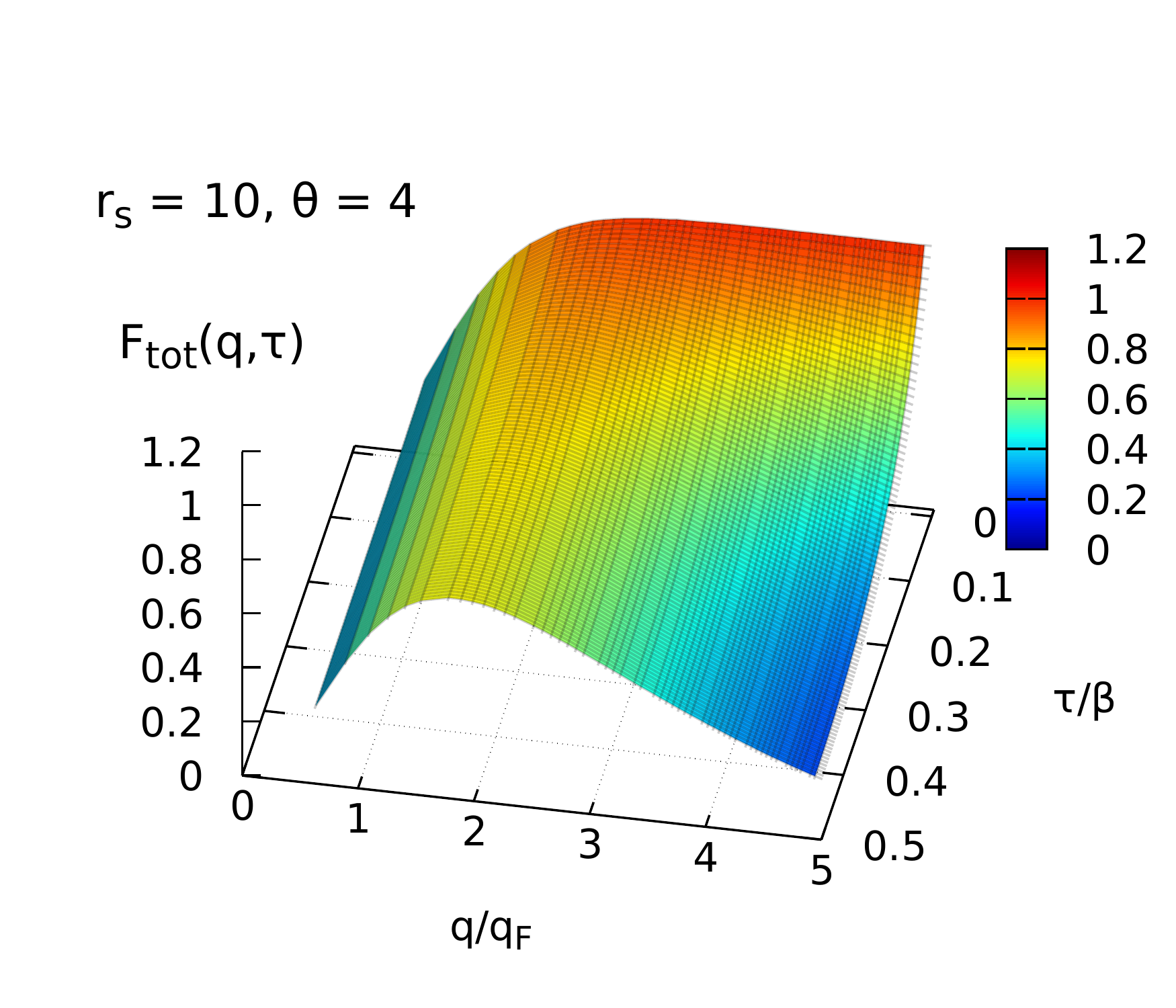}
\caption{\label{fig:UEG_ITCF_N34_rs10_theta}
\emph{Ab initio} PIMC results for the imaginary time density--density correlation function $F(\mathbf{q},\tau)$ in the $q$-$\tau$-plane. Shown are results for the unpolarized UEG with $N=34$ at $r_s=10$ and different values of the degeneracy temperature $\theta$.
}
\end{figure*}

Let us conclude this investigation of synthetic spectra by considering the impact of the temperature $\beta$. A corresponding analysis is shown in Fig.~\ref{fig:Gauss_beta} for three slightly different values of $\beta$. Indeed, the main difference between the spectra is given by the varying ratio of the peaks at positive and negative frequency. In stark contrast, we observe a pronounced influence of the temperature on $F(\tau)$ shown in the bottom panel of Fig.~\ref{fig:Gauss_beta}. From a mathematical perspective, $S(\mathbf{q},\omega)$ and $F(\mathbf{q},\tau)$ are completely equivalent representations of the same information. In practice, both domains tend to emphasise different aspects. For example, the peak width of the DSF is a concept of the $\omega$-domain and can be seen most clearly in $S(\omega)$, see Fig.~\ref{fig:Gauss_width} above. The $\tau$-domain, on the other hand, is intimately connected to the temperature of a system, which manifests as the somewhat subtle detailed balance relation Eq.~(\ref{eq:detailed_balance}) in the DSF. The corresponding symmetry [Eq.~(\ref{eq:symmetry})] of the ITCF, on the other hand, is substantially enhanced. 
From a physical perspective, this is not surprising, as the (inverse) temperature determines the characteristic variance of the imaginary-time diffusion process, cf.~Fig.~\ref{fig:Fig} above, and therefore decisively shapes the decay of correlations with $\tau$. In practice, it means that that $\tau$-domain constitutes the representation of choice for the extraction of the temperature from an XRTS measurement~\cite{Dornheim_T_2022}, see also the recent Ref.~\cite{Dornheim_T2_2022} for a more quantitative analysis. Lastly, we note that the utility of $F(\mathbf{q},\tau)$ has been confirmed by the independent re-analysis of an XRTS measurement of warm dense beryllium~\cite{DOPPNER2009182} by Sch\"orner \emph{et al.}~\cite{Schoerner_arxiv_2023}.

\subsection{Uniform electron gas\label{sec:UEG_results}}

\begin{figure*}\centering
\includegraphics[width=0.475\textwidth]{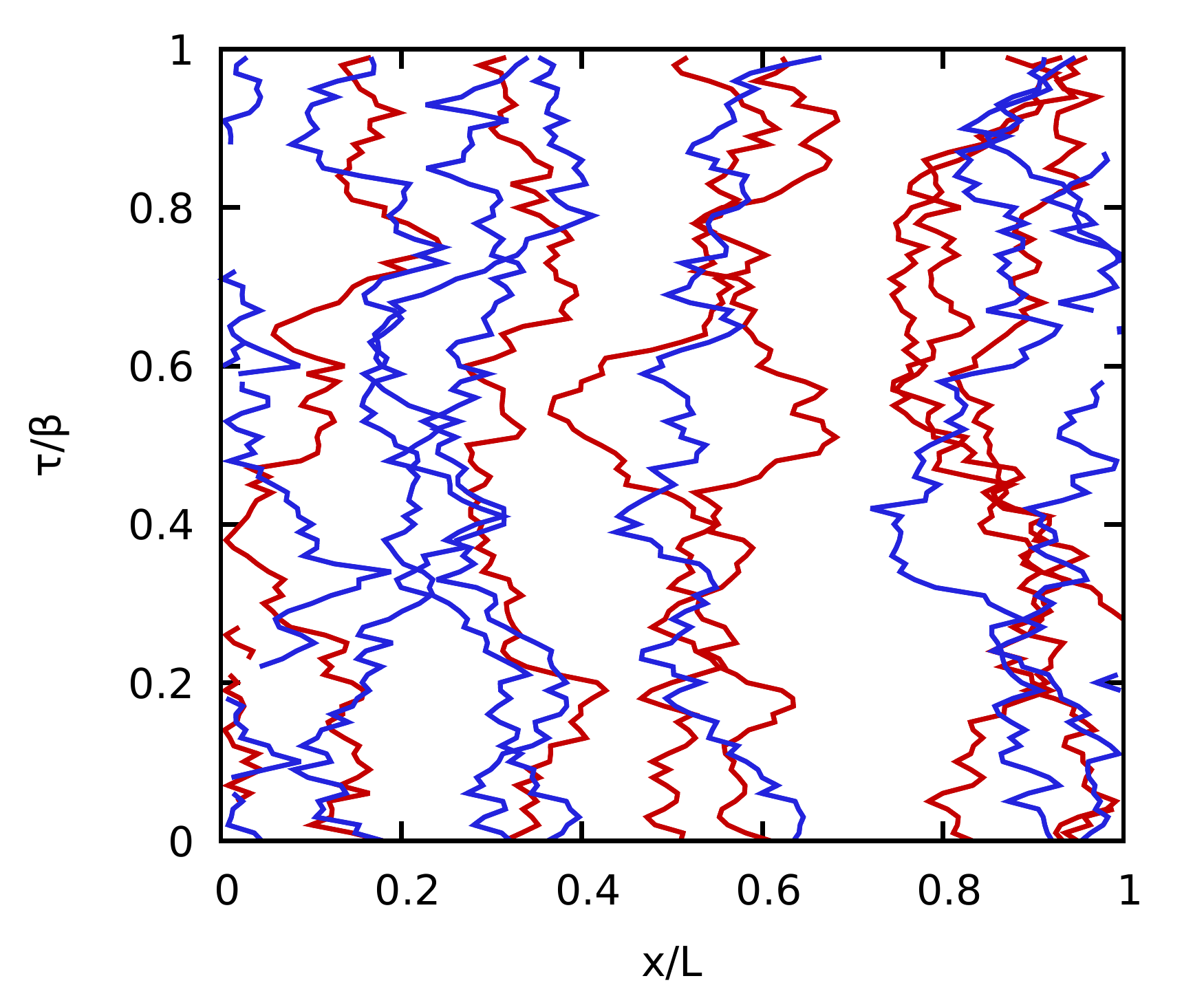}
\includegraphics[width=0.475\textwidth]{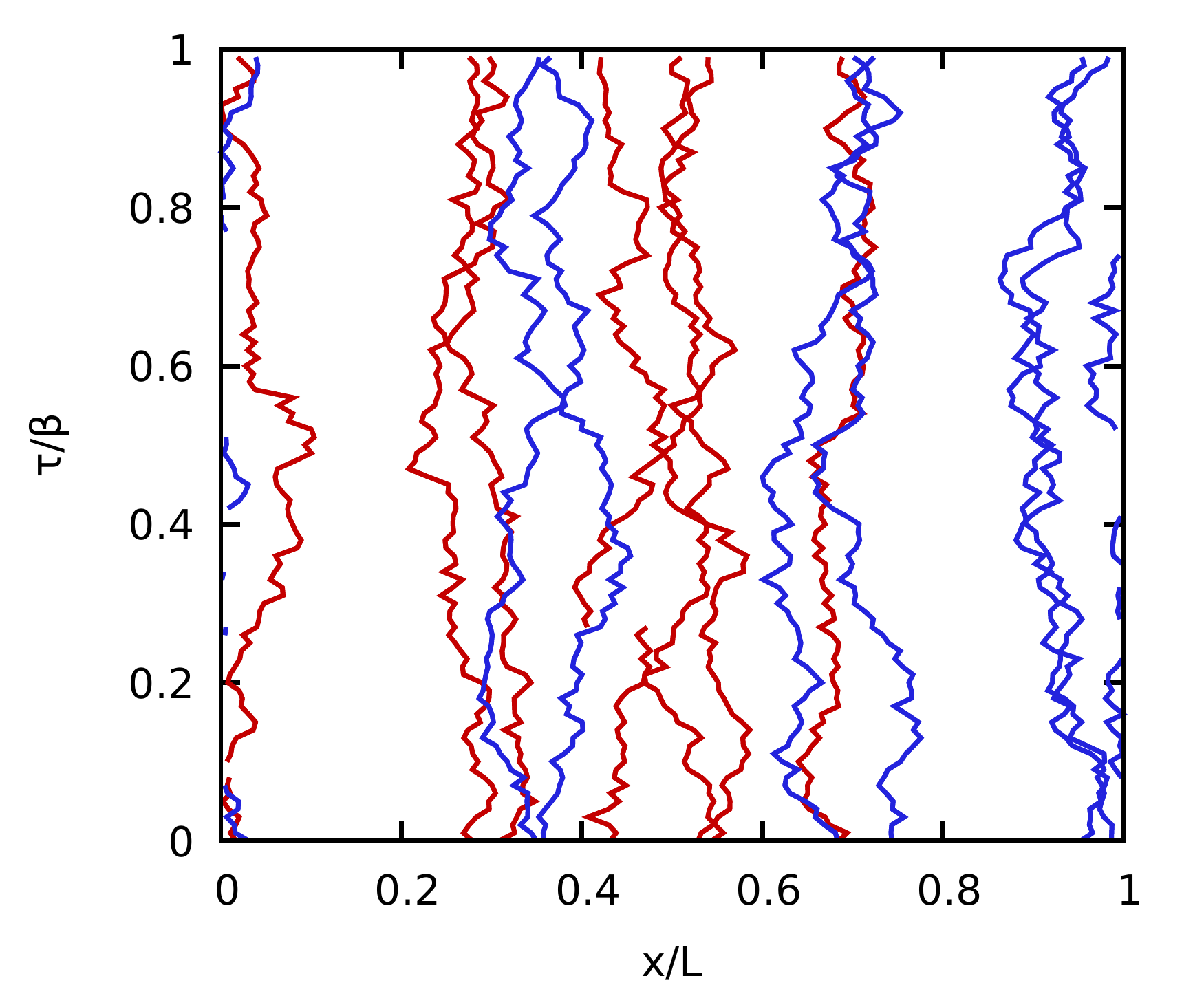}\\\vspace*{0.5cm}
\hspace*{0.04\textwidth}\includegraphics[width=0.35\textwidth]{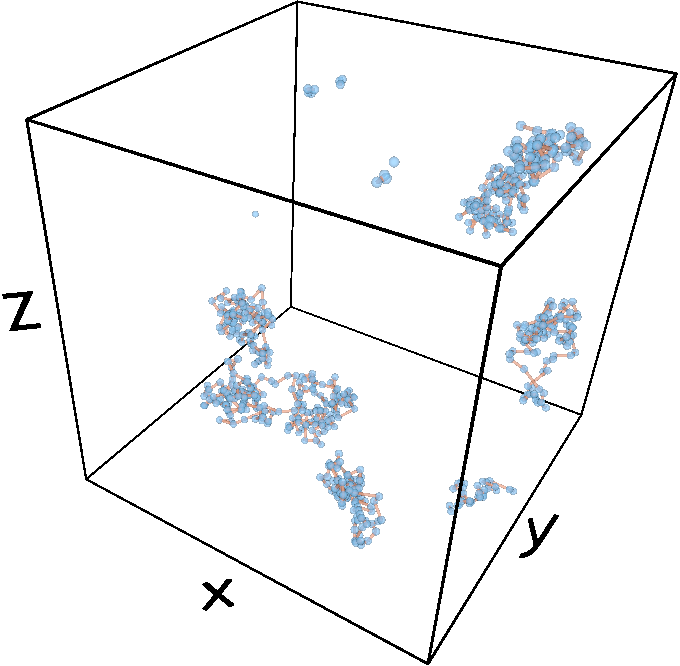}\hspace*{0.1\textwidth}\includegraphics[width=0.35\textwidth]{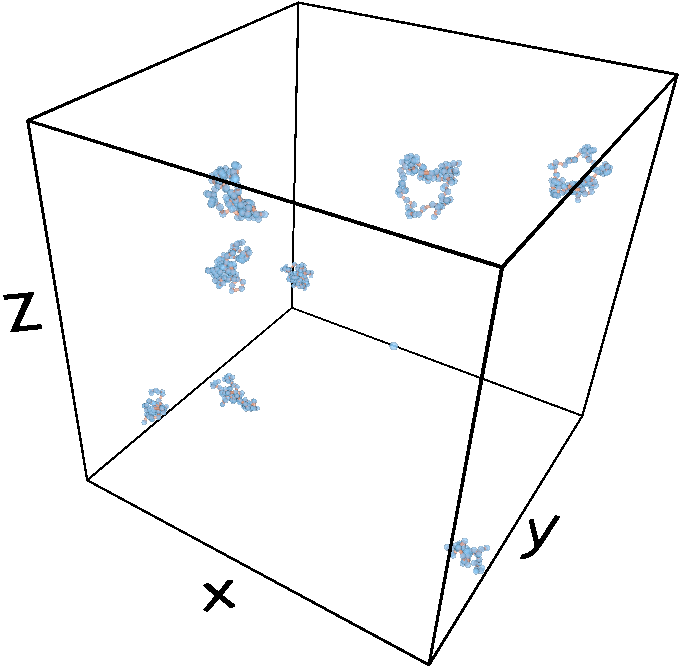}\\\vspace*{0.3cm}
\caption{\label{fig:Paths}
Snapshots of an \emph{ab initio} PIMC simulation of $N=14$ unpolarized electrons at $r_s=10$ for $\theta=1$ (left) and $\theta=4$ (right). Top row: paths of spin-up (red) and spin-down (blue) electrons in the $x$-$\tau$-plane. Bottom row: depiction of the configurations as paths in the $3D$ simulation cell.
}
\end{figure*}

Let us next turn our attention to exact simulation results for a physical system. In Fig.~\ref{fig:UEG_ITCF_N34_rs10_theta}, we show our \emph{ab initio} PIMC results for $F(\mathbf{q},\tau)$ for the UEG at $r_s=10$ and different values of the degeneracy temperature $\theta$. 
We note that it is sufficient to consider the interval $0\leq\tau\leq\beta/2$ due to the symmetry relation Eq.~(\ref{eq:symmetry}).
These parameters are located at the margin towards the strongly coupled electron liquid regime~\cite{dornheim_dynamic,dornheim_electron_liquid,quantum_theory}; still, we find almost no correlation induced features in the static structure factor $S(\mathbf{q})=F(\mathbf{q},0)$ for all depicted values of $\theta$. 
First and foremost, we observe a decay along the $\tau$-direction in the depicted interval for all values of the wave number $q$. This is expected, as correlations can only become weaker---or, in the extreme case, remain unaffected---due to the imaginary-time diffusion process that is sampled in our PIMC simulations.
In addition, it can clearly be seen that this decay of correlations is more pronounced at low temperatures.

This can directly be explained by recalling the discussion of the path sampling in Fig.~\ref{fig:Fig} above. In Fig.~\ref{fig:Paths}, we show snapshots of PIMC simulations of $N=14$ unpolarized electrons at $r_s=10$ for $\theta=1$ (left column) and $\theta=4$ (right column). More specifically, the top row shows path configurations (with the red and blue lines corresponding to spin-up and spin-down electrons) for the two different values of $\theta$. At the lower temperature, the paths exhibit a more pronounced diffusion along the $\tau$-direction compared to $\theta=4$. This is expected and a direct consequence of the definition of the thermal wave length $\lambda_\beta$ in Eq.~(\ref{eq:lambda}). Naturally, the larger displacements in coordinate space with increasing $\tau\leq\beta/2$ lead to a decreasing density--density correlation function.
For completeness, we also show the same configurations as paths in the $3D$ simulation box in the bottom row of Fig.~\ref{fig:Paths}. In this representation, the more pronounced imaginary-time diffusion process in the case of $\theta=1$ manifests as more extended paths. With increasing temperature, the paths become less extended, and attain the limit of classical point particles in the limit of $\beta\to0$.

\begin{figure}\centering
\includegraphics[width=0.475\textwidth]{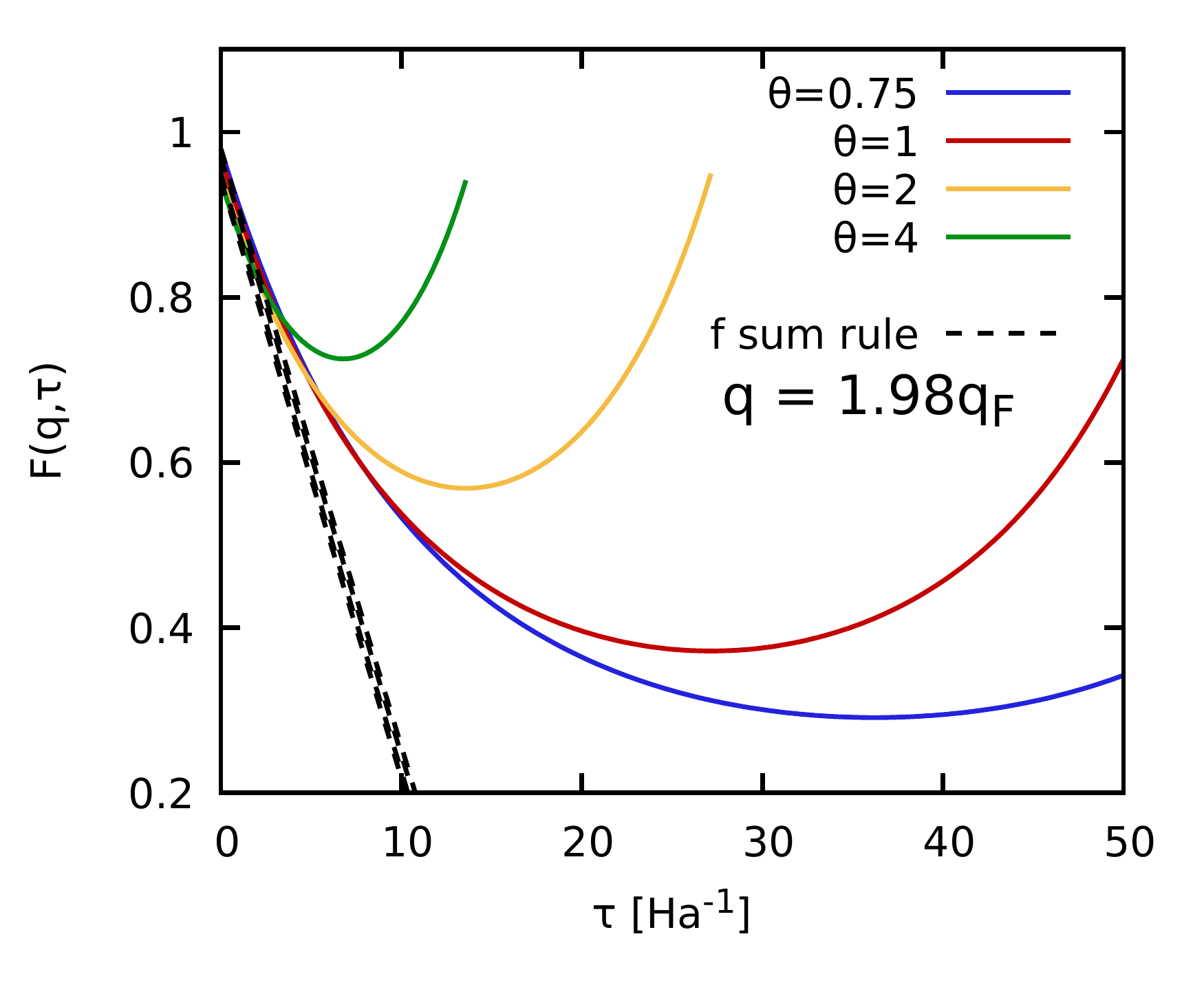}\\\vspace*{-1.cm}
\includegraphics[width=0.475\textwidth]{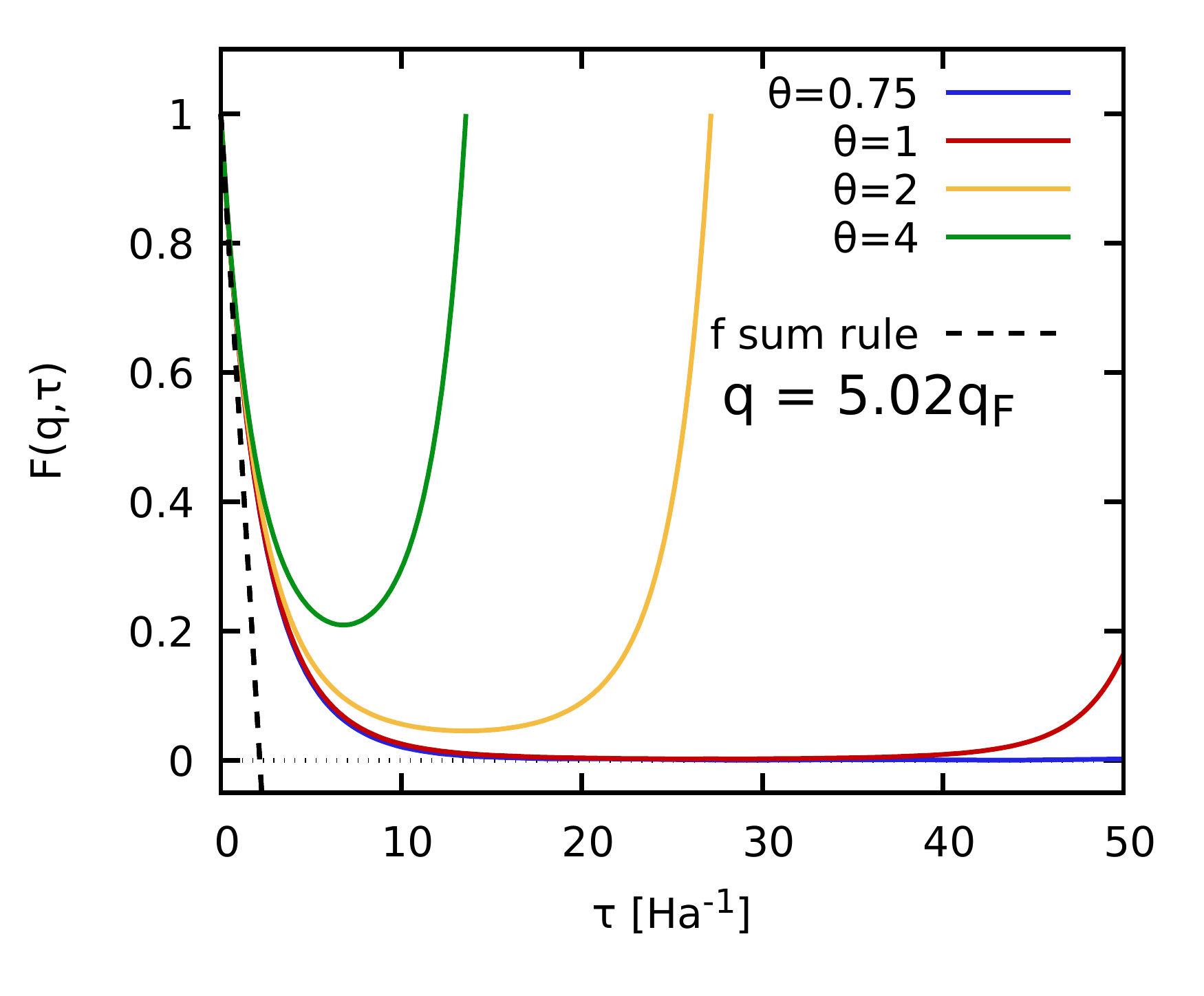}
\caption{\label{fig:ITCF_rs10_theta}
Imaginary-time dependence of $F(\mathbf{q},\tau)$ for two selected $q$-values taken from the full data set shown in Fig.~\ref{fig:UEG_ITCF_N34_rs10_theta}. The solid coloured curves correspond to the different valued of $\theta$, and the dashed black lines show a linear expansion evaluated from the exact f-sum rule, Eq.~(\ref{eq:f_sum_rule}). Note that we have not normalized the $\tau$-axis by the respective values of $\beta$.
}
\end{figure}

An additional trend that can be seen from Fig.~\ref{fig:UEG_ITCF_N34_rs10_theta} is that the imaginary-time decay of $F(\mathbf{q},\tau)$ becomes increasingly steep in the limit of large $q$.
This is investigated in more detail in Fig.~\ref{fig:ITCF_rs10_theta}, where we show $F(\mathbf{q},\tau)$ along the $\tau$-direction for two selected values of $q$. Specifically, the solid coloured curves correspond to the different values of $\theta$. First and foremost, we note that we have not re-scaled the $\tau$-axis by the respective values of the inverse temperature $\beta$ as in Fig.~\ref{fig:UEG_ITCF_N34_rs10_theta}. Therefore, the minima in the different curves are located at different positions in descending order of $\theta$. In fact, this representation gives a direct insight into the observed less pronounced decay of $F(\mathbf{q},\tau)$ along the $\tau$-direction: it is not the consequence of a less steep decay by itself, but rather of the reduced imaginary time that is available for the diffusion process.

This is further illustrated by the dashed black lines, which depict a linear expansion of $F(\mathbf{q},\tau)$ around $\tau=0$ evaluated from the exact f-sum rule, Eq.~(\ref{eq:f_sum_rule}). In particular,
the lines are parallel due to Eq.~(\ref{eq:f_sum_rule}), but the initial points are different due to the different static structure factors $S(\mathbf{q})=F(\mathbf{q},0)$.
In addition, Eqs.~(\ref{eq:f_sum_rule}) and (\ref{eq:moments_derivative}) clearly predict a parabolically increasing decay with respect to $q$ along the $\tau$-direction around $\tau=0$, which is reflected by the observed steep decay in the limit of large $q$ in Fig.~\ref{fig:UEG_ITCF_N34_rs10_theta}.

\begin{figure}\centering
\includegraphics[width=0.475\textwidth]{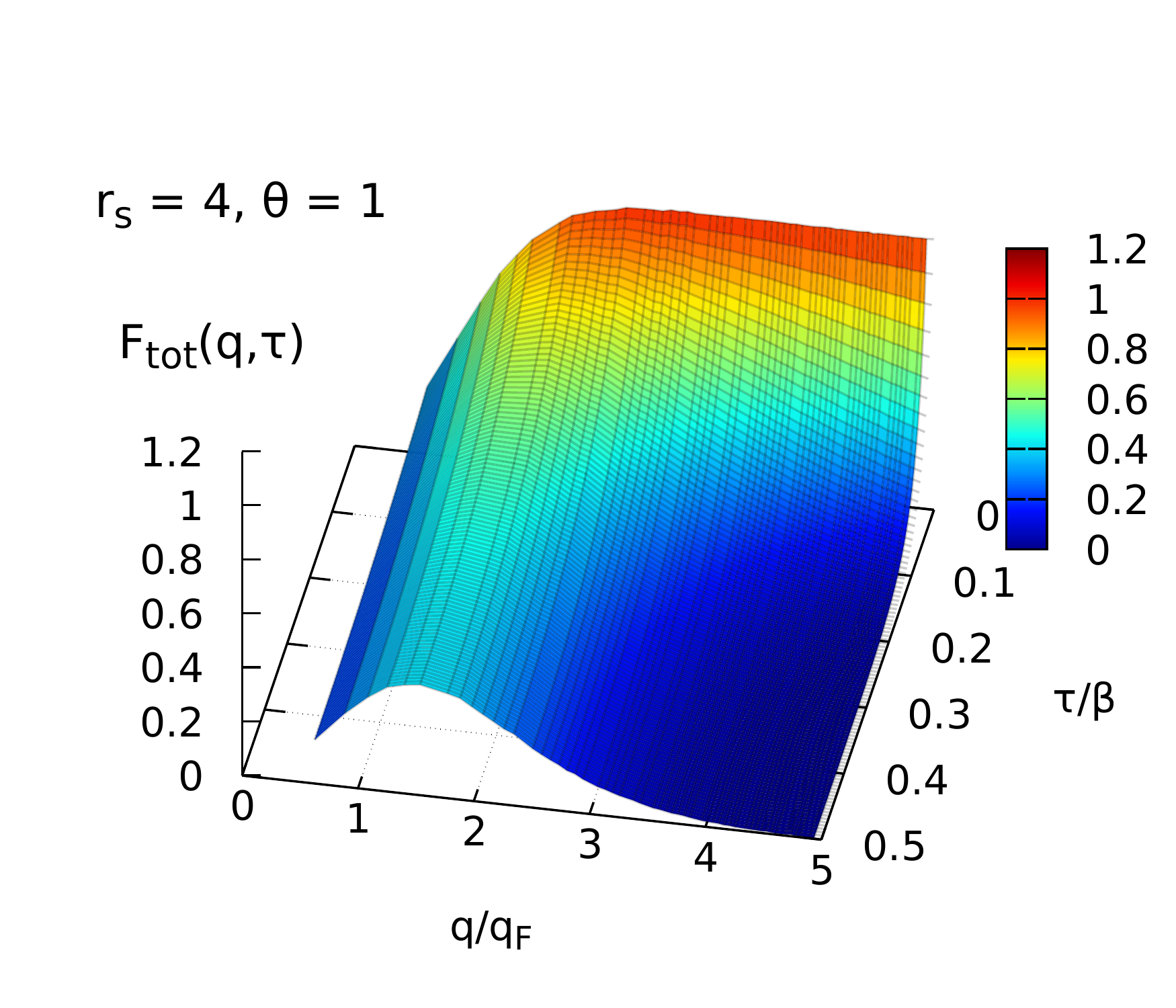}\\\vspace*{-1.6cm}
\hspace*{-0.5cm}\includegraphics[width=0.5\textwidth]{N34_rs10_theta1_3D_ITCF_total.pdf}\\\vspace*{-1.6cm}
\hspace*{-0.5cm}\includegraphics[width=0.5\textwidth]{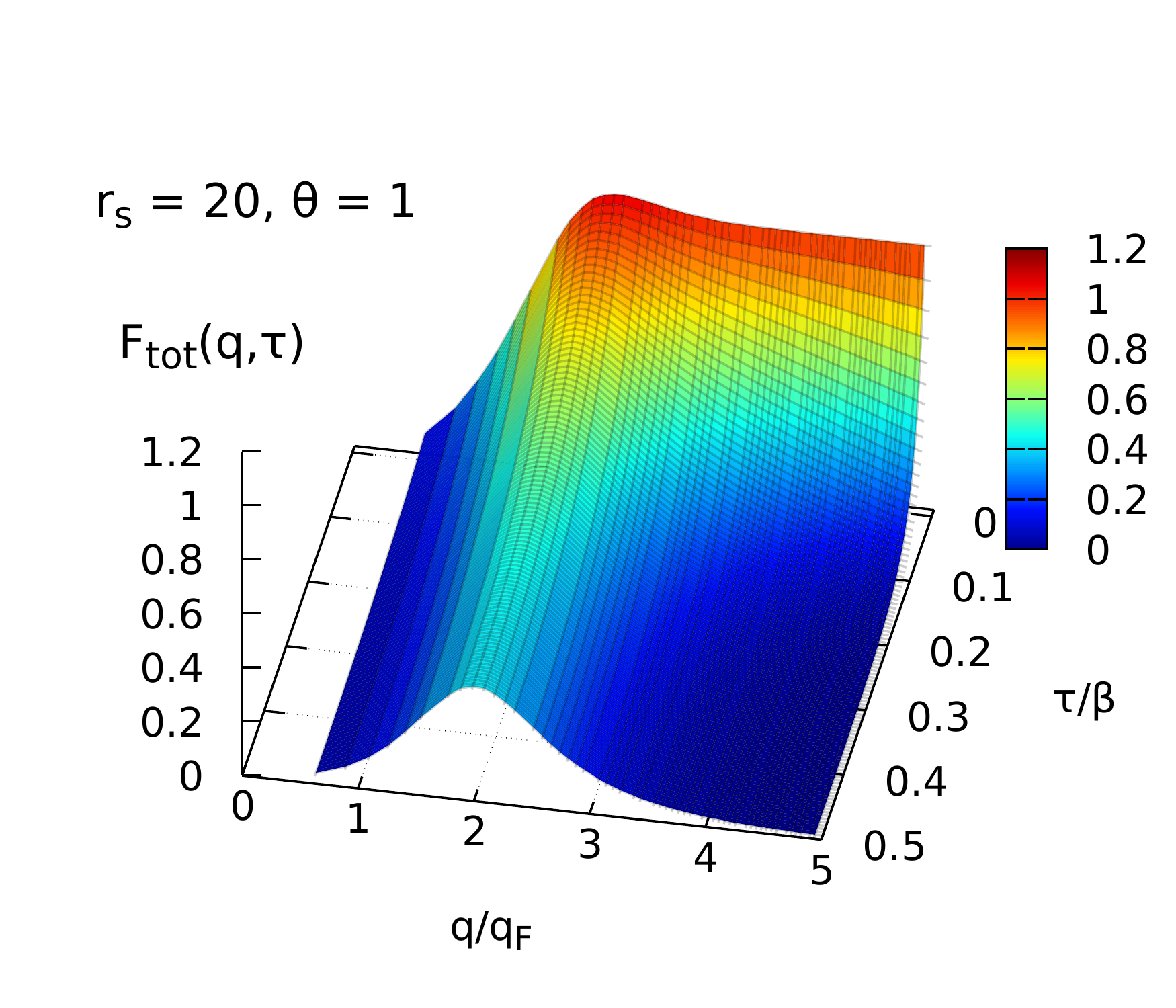}\vspace*{-0.6cm}
\caption{\label{fig:UEG_ITCF_N34_theta1_rs}
\emph{Ab initio} PIMC results for $F(\mathbf{q},\tau)$ in the $q$-$\tau$-plane. Shown are results for the unpolarized UEG with $N=34$ at $\theta=1$ and different values of the density parameter $r_s$.
}
\end{figure}

\begin{figure}\centering
\includegraphics[width=0.475\textwidth]{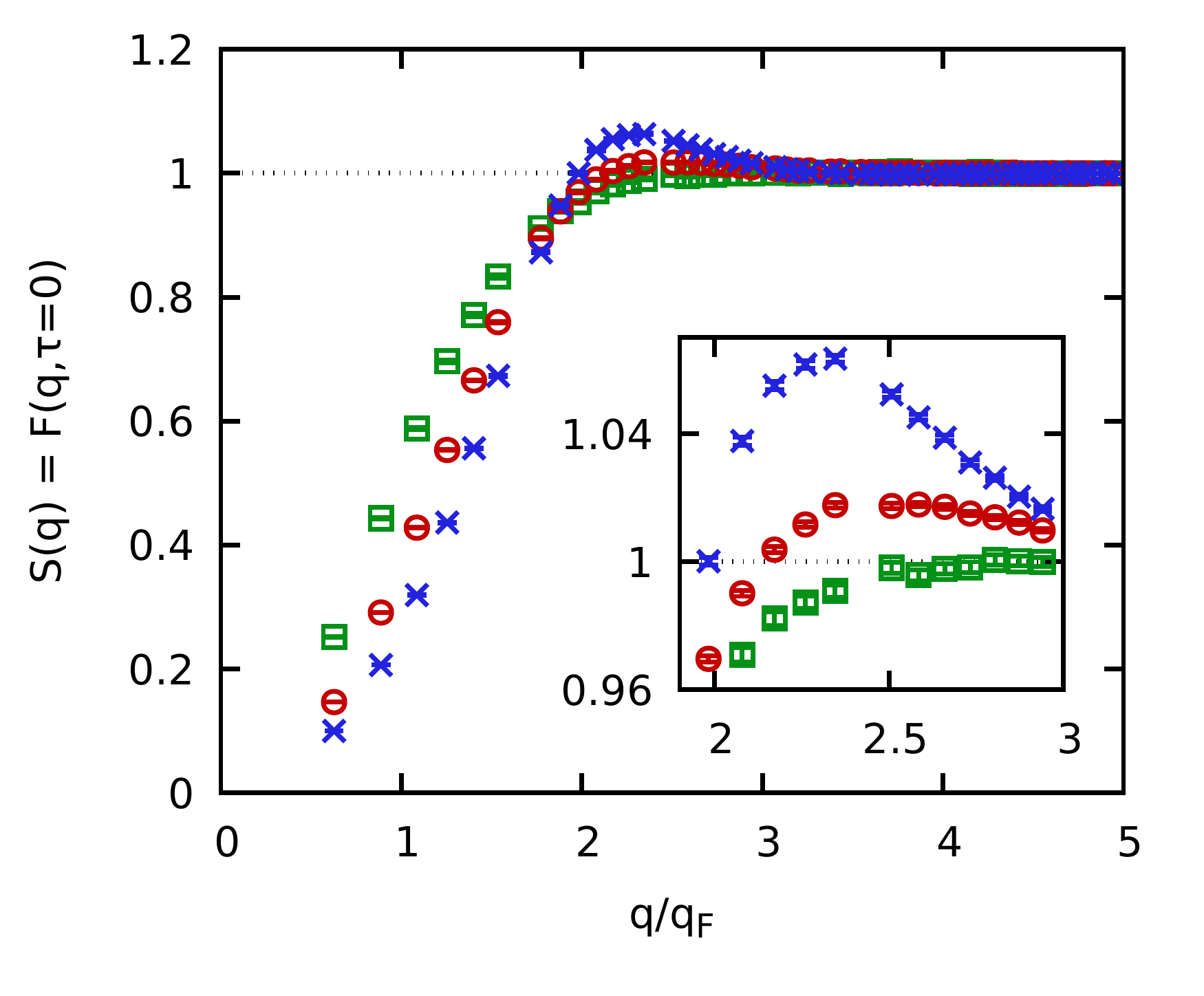}\\\vspace*{-1.34cm}
\includegraphics[width=0.475\textwidth]{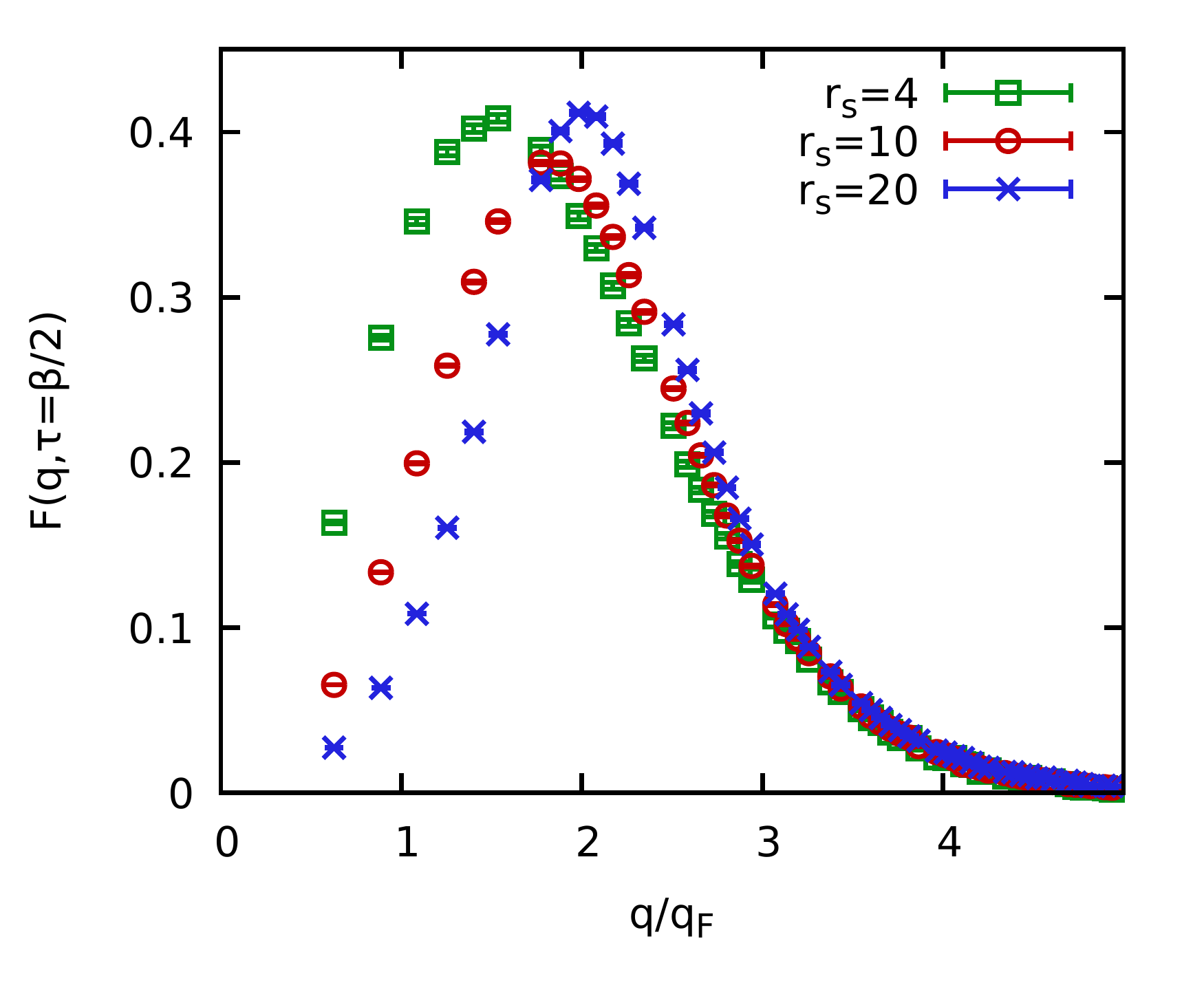}
\caption{\label{fig:Slice_theta1_rs}
Slices along the $q$-direction from the full ITCF shown in Fig.~\ref{fig:UEG_ITCF_N34_theta1_rs}. Top: static structure factor $S(q)=F(q,0)$. Bottom: $F(q,\beta/2)$.
}
\end{figure}

A further interesting research question is given by the dependence of $F(\mathbf{q},\tau)$ on the coupling strength. This is investigated in Fig.~\ref{fig:UEG_ITCF_N34_theta1_rs}, where we show our PIMC results for $F(\mathbf{q},\tau)$ at the electronic Fermi temperature $\theta=1$ for $r_s=4$ (top), $r_s=10$ (center), and $r_s=20$ (bottom).
In this case, the most pronounced trend is the increased structure in $S(\mathbf{q})$, i.e., in the limit of $\tau=0$.
This can be seen particularly well in the top panel of Fig.~\ref{fig:Slice_theta1_rs}, where we show this limit for all three considered values of $r_s$. The inset shows a magnified segment around the peaks at $r_s=20$ (blue)
and $r_s=10$ (red); no peak can be resolved for $r_s=4$ (green).

The bottom panel of Fig.~\ref{fig:Slice_theta1_rs} shows the same information for the \emph{thermal structure factor} $F(\mathbf{q},\beta/2$), where $F(\mathbf{q},\tau)$ attains its minimum value for all $q$. All three curves exhibit a qualitatively similar peak around $q=2q_\textnormal{F}$, which is indicative of a reduced decay along the $\tau$-direction. This, in turn, implies a shift in the spectral weight of the DSF $S(\mathbf{q},\omega)$ towards lower frequencies, cf.~the discussion of Fig.~\ref{fig:Gauss_w0} above. Interestingly, the peak height exhibits a non monotonic behaviour with respect to $r_s$, and is minimal for $r_s=10$. The peak position, on the other hand, is monotonically, although weakly, increasing with $r_s$.

\begin{figure}\centering
\includegraphics[width=0.475\textwidth]{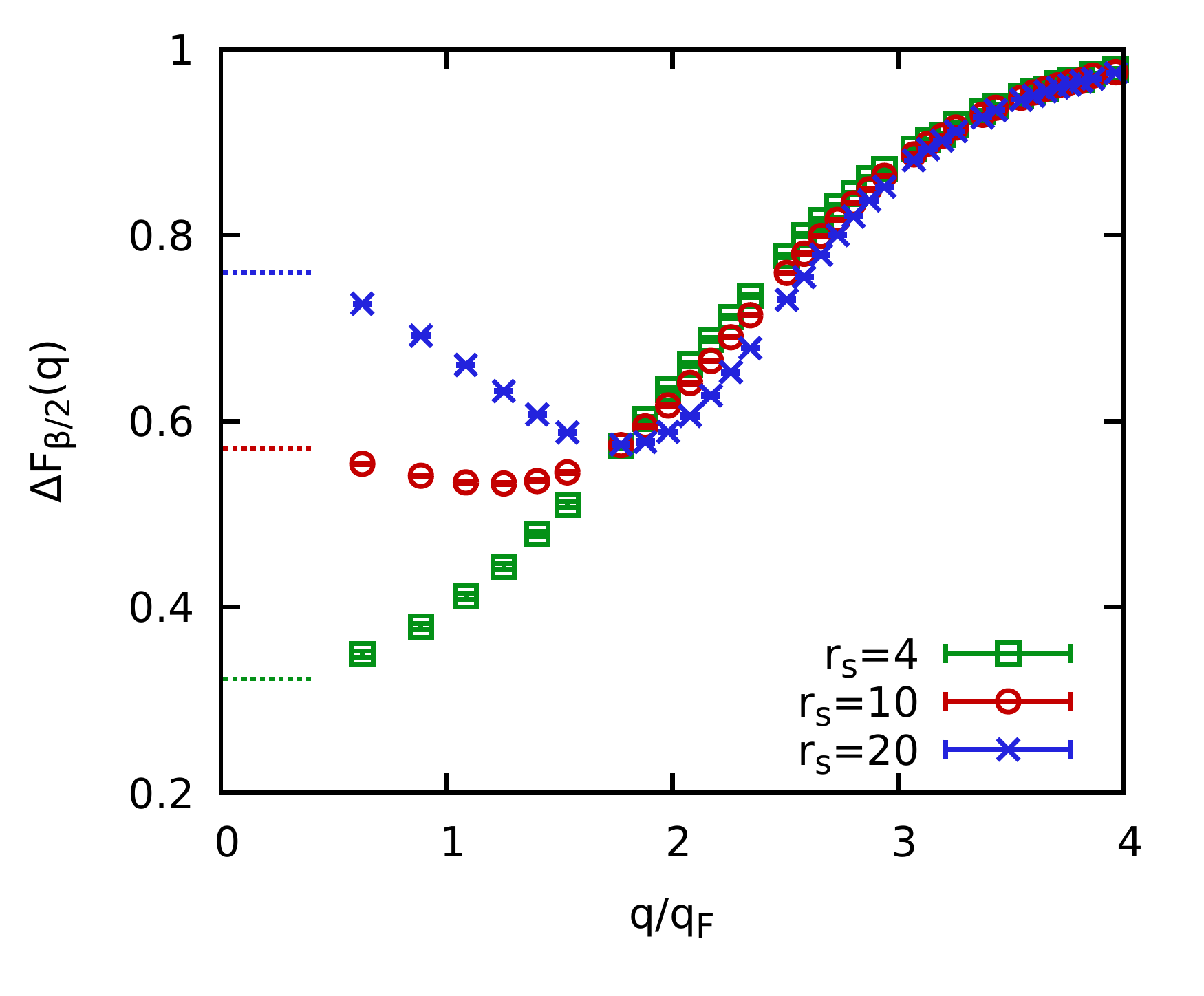}\\\vspace*{-1.34cm}
\includegraphics[width=0.475\textwidth]{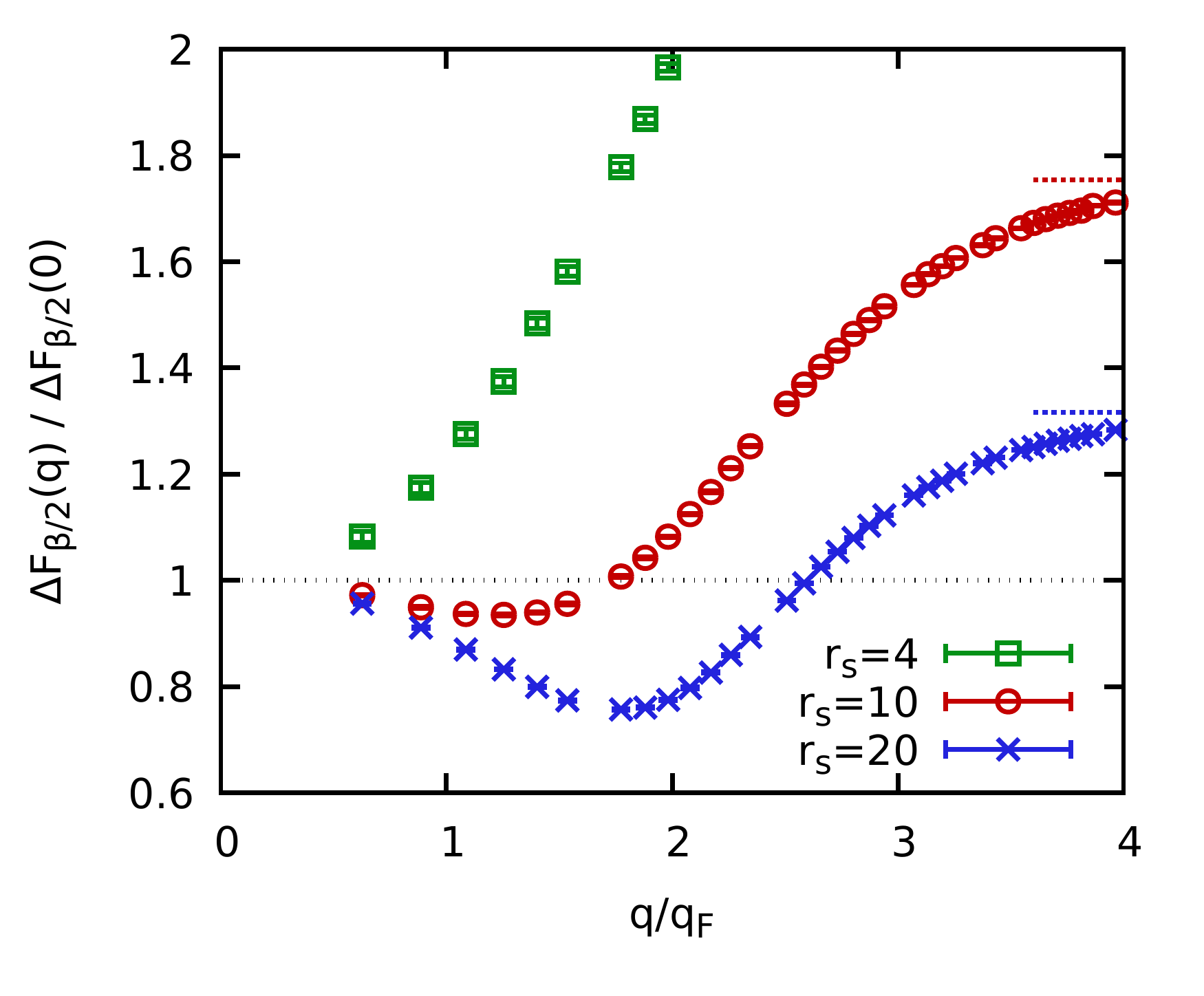}
\caption{\label{fig:Delta_Slice_theta1_rs}
Measure of relative $\tau$-decay [cf.~Eq.~(\ref{eq:decay_measure})] for the same conditions as in Fig.~\ref{fig:UEG_ITCF_N34_theta1_rs}. The vertical bars in the top panel indicate the $q\to0$ limit given by a sharp plasmon excitation at the plasma frequency, see Eq.~(\ref{eq:F_QP}). Bottom: same information, but normalized by the respective value at $q=0$, with the horizontal bars indicating the single-particle limit for $q\to\infty$.
}
\end{figure} 

To further investigate this effect, we define a measure of relative $\tau$-decay as
\begin{eqnarray}\label{eq:decay_measure}
\Delta F_\tau(q) = \frac{F(q,0)-F(q,\tau)}{F(q,0)}\ ,
\end{eqnarray}
where the values of zero and unity indicate no decay and full decay, respectively.
The results for $\Delta F_{\beta/2}(q)$ are shown in the top panel of Fig.~\ref{fig:Delta_Slice_theta1_rs}. Firstly, we see that all curves converge against each other for large $q$, and eventually attain the limit of $\lim_{q\to\infty}\Delta F_{\beta/2}(q)=1$. This is a direct consequence of the vanishing value of $F(q,\beta/2)$ in the single-particle limit. Around twice the Fermi wave number, the three curves in the top panel of Fig.~\ref{fig:Delta_Slice_theta1_rs} start to deviate from each other in a highly nontrivial way. The horizontal lines indicate the $q\to0$ limit of $\Delta F_{\beta/2}(q)$, which is determined by the sharp plasmon excitation at the plasma frequency $\omega_\textnormal{p}$, see also Eq.~(\ref{eq:F_QP}). For $r_s=4$, the curve is relatively featureless and smoothly interpolates between the $q=0$ and $q\to\infty$ limits. For $r_s=10$ and, in particular, $r_s=20$, the curves become non monotonic and exhibit a minimum at intermediate wave numbers. In other words, the decay along the $\tau$-direction is suppressed when the wave number is of the order of the average particle separation $d=2r_s$.

\begin{figure}\centering
\includegraphics[width=0.475\textwidth]{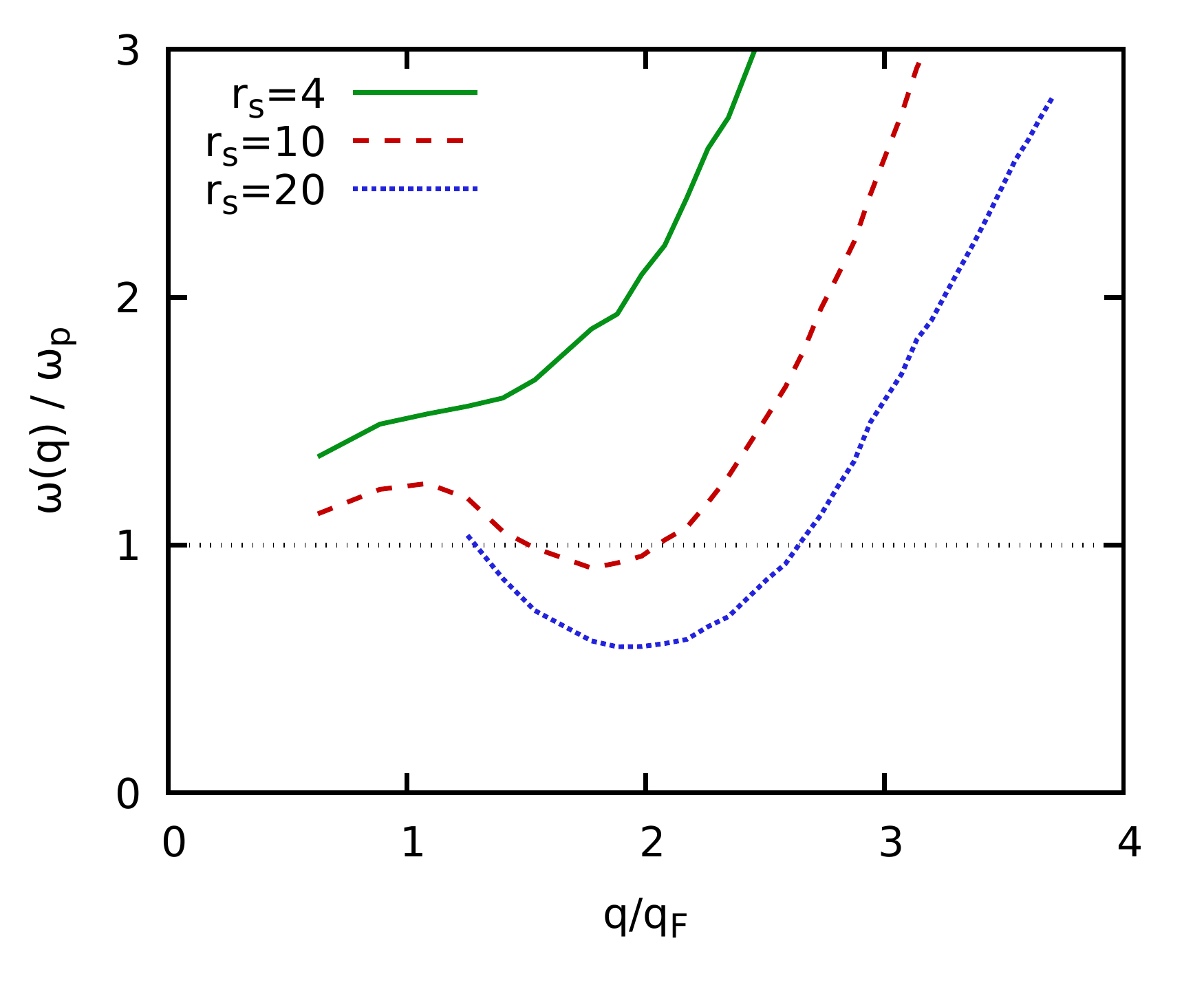}
\caption{\label{fig:Dispersion}
Position of the maximum in the DSF $\omega(q)$ for the same parameters as in Fig.~\ref{fig:UEG_ITCF_N34_theta1_rs}. Data taken from Ref.~\cite{dornheim_dynamic}.
}
\end{figure}

We have already mentioned several times that such a reduced decay implies a shift of spectral weight in the DSF towards lower frequencies $\omega$. This is indeed the case for the UEG at the present conditions; see the original Ref.~\cite{dornheim_dynamic} for all technical details on the corresponding calculations. The results for the wave number dependence of the position of the maximum in the DSF $\omega(q)$ are shown in Fig.~\ref{fig:Dispersion}. We observe a monotonic curve for $r_s=4$ and a distinct \emph{roton minimum} for $r_s=10$ and $r_s=20$ around the same position as the non monotonic behaviour of $\Delta F_{\beta/2}(q)$ in Fig.~\ref{fig:Delta_Slice_theta1_rs}. For completeness, we note that the corresponding energy reduction in the spectrum of density fluctuations was explained only recently by the alignment of pairs of electrons in Ref.~\cite{Dornheim_Nature_2022}, and further substantiated in the subsequent Ref.~\cite{Dornheim_Force_2022}.

To further illustrate the direct physical correspondence between the reduced $\tau$-decay, on the one hand, and the \emph{roton feature} in the DSF, on the other hand, we normalise $\Delta F_{\beta/2}(q)$ by its $q\to0$ limit in the bottom panel of Fig.~\ref{fig:Delta_Slice_theta1_rs}. We note that this is analogous to the normalisation with respect to the plasmon frequency of the dispersion relation $\omega(q)$ in Fig.~\ref{fig:Dispersion}. 
The resulting curves even more closely resemble the dispersion of the DSF for small to intermediate wave numbers, and exhibit the same qualitative trend. In particular, we find a comparable roton minimum for $r_s=10$ and $r_s=20$. This is unsurprising, as both $F(\mathbf{q},\tau)$ and $S(\mathbf{q},\omega)$ contain, by definition, the same physical information. Therefore, any physical process such as the \emph{roton feature} has to manifest itself both in the $\omega$- and in the $\tau$-domain. This is also evident from comparing Eqs. (\ref{eq:spectral}) and (\ref{eq:spectral_F}).



\section{Summary and Discussion\label{sec:summary}}

The present work is devoted to the investigation of the dynamic properties of correlated quantum many-body systems in the imaginary-time domain. In particular, we have argued that the usual approach in terms of the DSF $S(\mathbf{q},\omega)$ in frequency-space is not the only option as, by definition, the ITCF $F(\mathbf{q},\tau)$ contains exactly the same information, only in an unfamiliar representation. As such, some properties such as the peak width of a DSF are more easily accessible in the $\omega$-domain, whereas other physical effects such as the temperature or quantum mechanical delocalization are more clearly emphasised in $F(\mathbf{q},\tau)$. In fact, even nontrivial physical effects such as the \emph{roton feature} in the DSF~\cite{Dornheim_Nature_2022} can easily be identified in the ITCF.

Instead of attempting the notoriously difficult analytic continuation from QMC results for $F(\mathbf{q},\tau)$ to $S(\mathbf{q},\omega)$, we find that it can be highly advantageous to pursue the opposite direction. Indeed, transforming an experimental signal from the $\omega$-domain to the $\tau$-domain is straightforward and well-behaved with respect to the inevitable experimental noise~\cite{Dornheim_T_2022,Dornheim_T2_2022}. In addition, it allows for an easy deconvolution of the instrument function; this is an important point e.g.~in XRTS experiments~\cite{sheffield2010plasma,siegfried_review}. 
First and foremost, we have demonstrated that the availability of such experimental results for $F(\mathbf{q},\tau)$ directly allows for a number of physical insights. In particular, it allows for the model- and simulation-free extraction of important system parameters such as the temperature~\cite{Dornheim_T_2022,Dornheim_T2_2022} and quasi-particle excitation energies such as the plasmon frequency $\omega_\textnormal{p}$. 
In addition, experimental data for $F(\mathbf{q},\tau)$ can be straightforwardly compared to exact QMC results for the same quantity, which opens up the enticing possibility for unprecedented agreement between theory and experiment.

We are convinced that the proposed physical interpretation of the dynamic properties of correlated quantum systems in the $\tau$-domain will have a strong impact in a number of disciplines. A prime example is given the interpretation of XRTS experiments with WDM~\cite{falk_wdm}, which can be used for the systematic construction of model-free equation-of-state data bases~\cite{Falk_PRL_2014,Falk_HEDP_2012}. The latter are, in turn, of paramount importance for the description of astrophysical objects~\cite{drake2018high} and inertial confinement fusion applications~\cite{Betti2016}. 
In this context, we also note the advent of modern free-electron X-ray laser facilities such as the new European XFEL~\cite{Tschentscher_2017}, which, with their high repetition rate, will allow for high-quality results for $I(\mathbf{q},\omega)$ and, in this way, also $F(\mathbf{q},\tau)$; see the outlook of Ref.~\cite{Dornheim_review} for more details. This will facilitate measurements at a number of scattering angles and thus wave vectors, which will be important to resolve physical dispersion effects such as the \emph{roton feature}. 
Moreover, we stress that the basic idea to directly use $F(\mathbf{q},\tau)$ instead of $S(\mathbf{q},\omega)$ is in no way limited to either XRTS or the study of WDM, and can easily be applied to a wide range of other systems such as ultracold atoms~\cite{Filinov_PRA_2012,griffin1996bose,Godfrin2012,Dornheim_SciRep_2022}.

In addition to its considerable value for the interpretation of experiments, we note that the $\tau$-domain also opens up the way for new developments in the theoretical modelling of quantum many-body systems. For example, a central limitation of time-dependent density-functional theory~\cite{marques2012fundamentals}---either in its linear-response or full real-time formulation---is the absence of reliable external input such as the dynamic exchange--correlation kernel $K_\textnormal{xc}(\mathbf{q},\omega)$~\cite{Bohme_PRL_2022} beyond the local density approximation or generalized-gradient expansions~\cite{Moldabekov_PRL_2022}. While a number of model kernels exist either for the UEG~\cite{Panholzer_PRL_2018,Kaplan_PRB_2022} or more generic systems~\cite{Botti_PRB_2004,Ramakrishna_2020}, they are often restricted to the static limit. 
In this regard, exact QMC based input data for the imaginary-time dependence of density--density correlations can help in multiple ways~\cite{Dornheim_review}.
First, we propose to pursue \emph{imaginary-time} dependent DFT simulations of real materials, which can utilize exact, fully $\tau$-dependent information about exchange--correlation effects based on \emph{ab initio} QMC simulations. More specifically, one might either choose to combine a static, material-specific XC-kernel based on DFT calculations~\cite{Moldabekov_PRL_2022} with the $\tau$-dependence extracted from a PIMC simulation of either the UEG, or a more realistic system, or to directly attempt a generalization of suitable PIMC simulation data. While it is certainly true that some features about the dynamic density response of a given system of interest might be unresolvable in the $\tau$-domain (e.g. the width of a peak in the DSF), often one is primarily interested in frequency-integrated properties such as the electron--electron static structure factor $S_{\mathrm{ee}}(\mathbf{q})$ or its inverse Fourier transform, which is given by the electron-electron pair correlation function. To this end, operating in the $\tau$-domain does not constitute a disadvantage and might allow for the extraction of electron--electron correlations from DFT simulations with high fidelity~\cite{Dornheim_review}.

Finally, we note that the analysis of imaginary-time density--density correlation functions can be easily extended to higher-order correlators~\cite{Dornheim_JCP_ITCF_2021}.
These higher-order ITCFs can straightforwardly be estimated in PIMC simulations and will give new insights to dynamic three-body and four-body correlation functions~\cite{Dornheim_JPSJ_2021}. 
Moreover, such higher-order ITCFs are directly related to \emph{nonlinear response} properties~\cite{Dornheim_PRL_2020,Dornheim_PRR_2021,Moldabekov_JCTC_2022}. The latter are known to very sensitively depend on system parameters like the temperature and may thus constitute an additional tool of diagnostics~\cite{moldabekov2021thermal}.


\renewcommand{\theequation}{A\arabic{equation}}
\setcounter{equation}{0}
\appendix
\section*{Appendix: Imaginary-time version of the fluctuation–dissipation theorem}

\noindent Even though the imaginary-time fluctuation-dissipation theorem has found applications in the literature, to our knowledge, no derivation has ever been reported. The key lies in the selection of an appropriate imaginary-time integration interval that will allow the application of the Kramers-Kronig relation. 

Combination of the two-sided Laplace transform definition of the imaginary time intermediate scattering function, see Eq.~(\ref{eq:Laplace}), with the linear fluctuation-dissipation theorem, see Eq.~(\ref{eq:FDT}), leads to the expression
\begin{eqnarray}\label{eq:app1}
F(\boldsymbol{q},\tau)=-\frac{1}{\pi{n}}\int_{-\infty}^{+\infty}d\omega\frac{\Im\{\chi(\boldsymbol{q},\omega)\}}{1-e^{-\beta\omega}}e^{-\omega\tau}\,.
\end{eqnarray}
It is evident that integration of Eq.~(\ref{eq:app1}) within the imaginary time interval $\tau\in[0,\beta]$ exposes the $\omega=0$ pole,
\begin{eqnarray}\label{eq:app2}
\int_0^{\beta}d\tau\,{F}(\boldsymbol{q},\tau)=-\frac{1}{\pi{n}}\int_{-\infty}^{+\infty}d\omega\frac{\Im\{\chi(\boldsymbol{q},\omega)\}}{\omega}\,.
\end{eqnarray}
This opens up the way for the direct application of the Kramers–Kronig relation that simply yields
\begin{eqnarray}\label{eq:app3}
\int_0^{\beta}d\tau\,{F}(\boldsymbol{q},\tau)=-\frac{1}{{n}}\Re\{\chi(\boldsymbol{q},0)\}\,.
\end{eqnarray}
Given the fact that the static density response is a real quantity, one obtains the so-called imaginary-time version of the fluctuation-dissipation theorem,
\begin{eqnarray}\label{eq:app3}
\chi(\boldsymbol{q},0)=-n\int_0^{\beta}d\tau\,{F}(\boldsymbol{q},\tau)\,.
\end{eqnarray}

\section*{Acknowledgements}
This work was partially supported by the Center for Advanced Systems Understanding (CASUS) which is financed by Germany’s Federal Ministry of Education and Research (BMBF) and by the Saxon state government out of the State budget approved by the Saxon State Parliament.
The PIMC calculations were carried out at the Norddeutscher Verbund f\"ur Hoch- und H\"ochstleistungsrechnen (HLRN) under grant shp00026, and on a Bull Cluster at the Center for Information Services and High Performance Computing (ZIH) at Technische Universit\"at Dresden.

\bibliography{bibliography}
\end{document}